\documentclass[12pt,preprint]{aastex}

\shortauthors{Hinkle \& Lambert}
\shorttitle{HR 4049 Circumstellar Gas}

\begin{document}

\title{Infrared High-Resolution Spectroscopy of\\ 
Post-AGB Circumstellar Disks.\\
I. HR 4049 -- The Winnowing Flow Observed?}

\author{Kenneth H. Hinkle}
\affil{National Optical Astronomy Observatory\altaffilmark{1},\\
P.O. Box 26732, Tucson, AZ 85726-6732}
\email{hinkle@noao.edu}

\author{Sean D. Brittain\altaffilmark{2}}
\affil{National Optical Astronomy Observatory,\\
P.O. Box 26732, Tucson, AZ 85726-6732
\and 
Clemson University, Department of Physics and Astronomy,
Clemson, SC 29634; sbritt@clemson.edu}

\author{David L. Lambert}
\affil{The W.J. McDonald Observatory, University of Texas, Austin, TX 78712
USA; dll@anchor.as.utexas.edu}

\altaffiltext{1}{Operated by Association of Universities for Research in
Astronomy, Inc., under cooperative agreement with the National Science
Foundation}

\altaffiltext{2}{Michelson Fellow}

\begin{abstract}

High-resolution infrared spectroscopy in the 2.3-4.6 $\mu$m region
is reported for the peculiar A supergiant, single-lined spectroscopic
binary HR 4049.  Lines from the CO fundamental and first overtone,
OH fundamental, and several H$_2$O vibration-rotation transitions
have been observed in the near-infrared spectrum.  The spectrum of
HR 4049 appears principally in emission through the 3 and 4.6 $\mu$m
region and in absorption in the 2 $\mu$m region.  The 4.6 $\mu$m
spectrum shows a rich `forest' of emission lines.  All the spectral
lines observed in the 2.3-4.6 $\mu$m spectrum are shown to be
circumbinary in origin.  The presence of OH and H$_2$O lines confirm
the oxygen-rich nature of the circumbinary gas which is in contrast
to the previously detected carbon-rich material.  The emission and
absorption line profiles show that the circumbinary gas is located
in a thin, rotating layer near the dust disk.  The properties of
the dust and gas circumbinary disk and the spectroscopic orbit yield
masses for the individual stars, M$_{A~I}\sim$0.58 M$_\odot$ and
M$_{M~V}\sim$0.34 M$_\odot$.  Gas in the disk also has an outward
flow with a velocity of $> \atop \sim$1 km s$^{-1}$.  The severe
depletion of refractory elements but near-solar abundances of
volatile elements observed in HR 4049 results from abundance
winnowing.  The separation of the volatiles from the grains in the
disk and the subsequent accretion by the star are discussed.  Contrary
to prior reports, the HR 4049 carbon and oxygen isotopic abundances
are typical AGB values: $^{12}$C/$^{13}$C=6$^{+9}_{-4}$
and $^{16}$O/$^{17}$O$>$200.

\end{abstract}



\section{Introduction}

HR 4049 is the prototype for a class of peculiar, post-AGB,
single-lined, long-period, spectroscopic binaries
\citep{van_winckel_et_al_95}. The primary star of these binaries is an
early-type supergiant with very peculiar abundances.  Most objects in
this class exhibit strong infrared excesses of circumstellar origin
with carbon-rich circumstellar dust typically present.  The combination
of carbon-rich material, high luminosity, and location out of the
galactic plane forms the basis for the post-AGB designation.  The
peculiar designation stems from a photospheric abundance pattern
characterized by a severe deficiency of refractory (high dust
condensation temperature) elements and a near-solar abundance for
volatile (low dust condensation temperature) elements.  The abundance
anomalies indicate that the present photosphere contains material from
which refractory elements have been very largely removed, i.e., a
winnowing of dust from gas has occurred.

The basic characteristics of the prototype object HR 4049 are well
known.  The photosphere has an effective temperature of about 7500
K and shows extreme abundance differences between refractory and
volatile elements: for example,  HR 4049 has [Fe/H]$\sim$-4.8  but
[S/H]$\sim$-0.2 \citep{waelkens_et_al_91b,takada_hidai_90}.  The
orbital period of HR 4049 is 430 days \citep{waelkens_et_al_91a},
leading to a minimum separation between the two stars of 190
R$_\odot$.  \citet{bakker_et_al_1998} pointed out that the orbit
requires a phase of common envelope evolution when the primary was
at the tip of the AGB and had a radius of $\sim$ 250 R$_\odot$.
This phase altered the masses and abundances of the components.

The HR 4049 infrared excess is pronounced redward of $\sim$1 $\mu$m and is
very well fit by a single blackbody at a temperature of about 1150
K and attributed to radiation from the tall inner walls of an optically
thick circumbinary disk \citep{dominik_et_al_2003}.  
A circumbinary Keplerian rotating disk appears a common feature of the
HR 4049 class of post-AGB binaries \citep{de_ruyter_et_al_06}.
The inner walls of the HR 4049 circumbinary disk 
are $\sim$ 10 AU from the binary or 50 times the radius of the
supergiant.  Our line of sight to the binary nearly grazes the edge
of the disk; the angle of inclination of the line of sight to the
normal to the disk is about 60$^{\circ}$.  A cartoon of the system
is shown in Figure 1.

Superimposed on the infrared dust continuum are emission features.
\citet{waters_et_al_1989} detected features due to polycyclic aromatic
hydrocarbons (PAHs).  This result was confirmed by the ISO/SWS spectrum
of \citet{beintema_et_al_1996}.  \citet{geballe_et_al_1989} confirmed a
C-rich circumstellar environment by detecting 3.43 and 3.53
$\mu$m emission features later identified with hydrogen-terminated
crystalline facets of diamond \citep{guillois_et_al_1999}.  Remarkably,
\citet{dominik_et_al_2003} report that the gas molecular species seen
in the infrared (ISO) spectrum are those expected of an O-rich mixture,
suggesting that the HR 4049 circumbinary environment is a blend of
C-rich dust and O-rich gas.

Optical spectroscopy offers some information on the disk's gas.
\citet{bakker_et_al_1998} report changes in the H$\alpha$ line profile
with orbital phase.  \citet{bakker_et_al_1998} and
\citet{bakker_et_al_1996} detected a broad ($\sim$15 km s$^{-1}$)
stationary emission component in the Na I D$_2$ and [O I] 6300
\AA\ lines which they attributed to the circumbinary disk.  However,
the infrared offers much more readily interpreted signatures of gas in
and around the binary. \citet{lambert_et_al_1988} detected the
$^{12}$C$^{16}$O first overtone spectrum in absorption with an
excitation temperature of about 300 K.  \citet{cami_yamamura_2001} on
analyzing an ISO/SWS spectrum of CO$_2$ emission bands found strong
contributions from isotopomers containing $^{17}$O and $^{18}$O which
they interpreted as $^{16}$O/$^{17}$O = 8.3 $\pm$ 2.3 and
$^{16}$O/$^{18}$O = 6.9$\pm$0.9.

The origins of the HR 4049 class of chemically peculiar 
supergiants and the structure
of their circumbinary/circumstellar material remain ill-understood. New
observational attacks appear to be essential.  In this paper we report
on a detailed look at several regions of the
2-5 $\mu$m infrared spectrum of HR 4049. 

\section{Observations \& Data Reduction}

The spectrum HR 4049 was observed at high resolution at a number of
near-infrared wavelengths in the 2.3 to 4.6 $\mu$m region using the
8\,m Gemini South telescope and the NOAO high-resolution near-infrared
Phoenix spectrometer
\citep{hinkle_et_al_98,hinkle_et_al_00,hinkle_et_al_03}.  Phoenix is a
cryogenically cooled echelle spectrograph that uses order separating
filters to isolate sections of individual echelle orders.  The detector
is a 1024$\times$1024 InSb Aladdin II array.  Phoenix is not cross dispersed
and the size of the detector in the dispersion direction limits the
wavelength coverage in a single exposure to about 0.5\%, i.e. 1550 km
s$^{-1}$, which is 0.012 $\mu$m at 2.3 $\mu$m (22 cm$^{-1}$ at 4300
cm$^{-1}$) and 0.024 $\mu$m at 4.6 $\mu$m (11
cm$^{-1}$ at 2100 cm$^{-1}$).  One edge of the detector is
blemished so the wavelength coverage is typically trimmed a few percent
to avoid this area.  Wavelength coverage is limited overall to 0.9-5.5
$\mu$m  by the InSb detector material.  All the spectra discussed here 
were observed with the widest (0.35 arcsecond) slit resulting in a
spectral resolution of R=$\lambda$/$\Delta$$\lambda$= 50,000.  The
central wavelengths of the regions observed are listed in Table 1.

The thermal brilliance of the sky makes observations longward of
$\sim$4 $\mu$m much more difficult than in the non-thermal 1-2.4$\mu$m
region.   However, for a bright star like HR 4049 this only slightly
increases the already short integration time.  Thermal infrared
observations were done using standard infrared observing techniques
\citep{joyce_92}.  Each observation consists of multiple integrations
at several different positions along the slit, typically separated by
4\arcsec~on the sky.  At thermal infrared wavelengths the telluric
lines are in emission.  In order not to saturate the telluric emission
lines the limiting exposure time is about 30 seconds at 4.6 $\mu$m.  For
longer integration times, multiple exposures can be coadded in the
array controller to make up a single exposure.  However, HR 4049 is so
bright that total exposure times of only 10 to 20 seconds were
required.  The delivered image FWHM at the spectrograph varied from
0.25\arcsec -- 0.80\arcsec during the nights that spectra were taken.
With the positions along the slit separated by several arcseconds the
resulting spectral images were well separated on the detector.

An average flat observation minus an average dark observation was
divided into each frame observed and frames with the star at different
places along the slit were then differenced and the spectrum extracted
using standard IRAF\footnote{The IRAF software is distributed by the
National Optical Astronomy Observatories under contract with the
National Science Foundation.} routines.  A hot star, with no intrinsic
spectral lines in the regions observed, was also observed at each
wavelength setting.  The hot star was observed at airmass near that of
HR 4049 and the HR 4049 spectrum was later divided by the hot star
spectrum to ratio the telluric spectrum from the HR 4049 star
spectrum.  Wavelength calibrations were computed by using a set of
telluric wavelengths obtained from the hot star spectra.  The
wavelength calibration yielded residuals of typically
0.25 km s$^{-1}$.

Observations of HR 4049 were taken in the 2 and 3 $\mu$m region as
well as the 4.6 $\mu$m region.
For a bright star the stellar signal is much stronger than background
radiation in these spectral regions and as a result the observations
are much less challenging than 4.6 $\mu$m observations.  The Phoenix
observing technique for this spectral region has previously been
described in \citet{smith02}.

\section{Analysis of the Spectra}

Observations of the 2.3, 3.0, and 4.6 $\mu$m regions reveal lines from
just three molecular species: CO, OH, and H$_2$O.  As discussed in \S1,
HR 4049 has a carbon-rich circumstellar envelope but we did not
identify any molecules associated with conditions where C$>$O.  In the
4.6 $\mu$m region we searched for C$_3$ and CN which should be
prominent if C$>$O.  Rather, we report on the new detection of a rich
4.6 $\mu$m forest of CO fundamental and H$_2$O lines.  The 4.6 $\mu$m
region atomic hydrogen lines Pfund $\beta$ and Humphreys $\epsilon$, if
present, are blended with H$_2$O lines.  The 4.6 $\mu$m HR 4049 emission 
line spectrum is very rich making the identification
of occasional atomic features problematic.  The circumstellar continuum
at 4.6 $\mu$m is approximately 25 times more intense than the continuum of 
the supergiant \citep{dominik_et_al_2003} so features of stellar origin
will be highly veiled.  At 3 $\mu$m we make a first detection of the OH
fundamental vibration-rotation lines in HR 4049.  Before exploring
the emission line spectrum, we revisit in \S3.1 the
2.3 $\mu$m CO first overtone spectrum detected previously by
\citet{lambert_et_al_1988}.  

All velocities in this paper are
heliocentric.  In order to compare heliocentric velocities of HR 4049
with microwave observations add -11.6 km s$^{-1}$ to convert to the
local standard of rest.

\subsection{CO First Overtone}

Our observations confirm and extend the discovery of first-overtone
($\Delta$v=2) vibration-rotation CO lines in absorption
\citep{lambert_et_al_1988}. At the 2.3 $\mu$m wavelength of the CO
first overtone, the continuum from the dust is about four times
that of the supergiant.  Thus, the CO absorption lines
should be formed along the lines of sight to the circumbinary disk.  By
inspection (Figure 2), it is apparent that the rotational and
vibrational temperatures are low; high rotational lines of the 2-0
band are absent and the 3-1 (and higher) bands are weak or
absent relative to the low rotational lines of the 2-0 band. All
of the prominent lines are attributable to the most common isotopic
variant, $^{12}$C$^{16}$O, but weak 2-0 $^{13}$CO lines are
detectable\footnote{We follow the convention of omitting the
superscript mass number for the most common isotope.  Hence
$^{12}$C$^{16}$O appears as CO, etc.}.

The first set of observations from February 2002 show the CO lines
at a radial velocity of -33$\pm$0.5 km s$^{-1}$.  The lines reach
maximum strength at J''$\sim$7, suggesting a rotational excitation
temperature of $\sim$300 K.  At J''=7, the R branch lines are about
17\% deep and are resolved with a FWHM of 16 km s$^{-1}$ compared
to the instrumental resolution of 6 km s$^{-1}$.  Weak $^{13}$CO
lines are detected with depths for the R18 to R23 lines of about
4\%.  Comparing lines of similar excitation suggests that
$^{12}$C/$^{13}$C$\sim$10.  The observed regions cover the strongest
predicted C$^{17}$O lines (2-0 lines near J''$\sim$ 7) but these lines can
not be convincingly identified in the spectra (Figure 2) demanding
C$^{16}$O/C$^{17}$O $> \atop \sim$100.  A yet more stringent limit
can be applied (\S4) by modeling.

After the original observations, additional data were collected to
extend the excitation range of the lines.  For instance, observations
made in December 2002 included higher J lines than observed previously.
Observations in December 2005 covered the low J P branch required
for curve-of-growth analysis.  Some wavelength intervals
were reobserved over the 2002 -- 2005 interval to check for
variability.  The radial velocity was in all cases unchanged from
the -33 km s$^{-1}$ measured in the original data set.  This velocity
is nearly equal to the -32.09 $\pm$0.13 km s$^{-1}$ systemic velocity
of the spectroscopic binary \citep{bakker_et_al_1998}.

The CO first overtone line profiles are symmetrical with no hint of an
emission component.  The line strengths showed no temporal
variability.  In fact, the line intensities are quite similar to those
reported by \citet{lambert_et_al_1988}.  Similarly, the velocity is
identical to the earlier reported value. The profile of the lowest
excitation line, 2-0 R0, differs from others in that it possibly has a
weak blue-shifted component.  However, the 2-0 R0 line lies in a region
with a complex telluric spectrum which is difficult to ratio out of the
HR 4049 spectrum.  In the February 2003 spectrum, the 2-0 R0 line
appears to have components at -23.6 and -32.9 km s$^{-1}$. On other
dates, the blue-shifted component is less clearly resolved suggesting
that it is either of variable velocity, affected by overlying emission
of variable intensity and/or velocity, or a relic of the reduction
process.  If this blue-shifted component exists, it must originate in
very cold gas (T ${<}\atop{^\sim}$ 5 K) because the component is not
detectable in the 2-0 R1 line.

\subsection{CO Fundamental}

In sharp contrast to the spectrum at 2.3 $\mu$m where a sparse
collection of weak absorption lines are found, the spectrum near
4.6 $\mu$m is rich in emission lines (Figure 3).  Emission lines
were identified from four isotopic variants:  CO, $^{13}$CO,
C$^{17}$O, and C$^{18}$O with roughly equal intensities for all
variants. The rarer isotopic variants $^{13}$C$^{17}$O, $^{13}$C$^{18}$O,
and $^{14}$CO were searched for but are not present.  Absorption
below the local continuum is also seen in the profiles of the lowest
excitation 1-0 CO lines in this interval.  The observed spectral
interval provides lines mostly from the 1-0 and 2-1 CO bands but a
few R branch lines of the 3-2 CO band are clearly present. The
maximum observable rotational level, J''$\sim$30, is similar to that seen in
the CO first overtone.  Table 2 lists the detected fundamental lines
of the four CO isotopic forms.

Blending with other lines of different CO isotopes, vibration-rotation
transitions, or H$_2$O lines is common and results in an apparent
variety of profiles.  All unblended emission lines are double-peaked
-- see Figures 4 -- 8.  For all but the lowest excitation lines the
blue and red peaks are of similar intensity, but characteristically
with the blue peak slightly weaker than the red, and occur at
velocities of -38.9 $\pm$0.4 and -28.4 $\pm$ 0.5 km s$^{-1}$,
respectively.  The central valley of the emission profile has a
velocity of -33.7 $\pm$ 0.3 km s$^{-1}$.  Velocities are not dependent
on the isotopic species.  The observed, i.e. uncorrected for
instrumental profile, full-width at zero intensity (FWZI) of the
emission profile for the weaker lines is $\sim$ 27 km s$^{-1}$,
with stronger lines having FWZI up to $\sim$ 35 km s$^{-1}$.  Due
to the high line density, the FWZI is a difficult parameter to
measure and our values carry an uncertainty of several km s$^{-1}$.
The observed full-width at half maximum (FWHM) similarly depends
on the line strength but much less dramatically than the FWZI.
Typical FWHM values are $\sim$ 19 km s$^{-1}$.

At the observed resolution of $\lambda$/$\Delta\lambda$=50000, the
instrument profile has a significant impact on the observed line
profiles and FWZI.  Assuming a Gaussian instrumental profile equal to
the 6 km s$^{-1}$ FWHM spectral resolution,
deconvolution of this instrumental profile from the observed line
profile gives a true FWZI of 18 km s$^{-1}$.  The line profiles are
strongly smoothed by the instrumental profile.  The observations
show blue and red sides of the CO lines rising $\sim$20\% above the
body of typical CO emission lines with the peak at each edge having a
FWHM of $\sim$ 4 km s$^{-1}$.  For more strongly saturated lines, e.g.
low excitation $^{13}$CO lines the FWHM of the blue and red emission
spikes are $\sim$8 km s$^{-1}$.  The intrinsic profile clearly has much
stronger emission peaks.

The combined absorption-emission profile for the 1-0 CO lines is shown
best by the R2 and R5 lines (Figure 9). After allowance for similar
blends, the profiles of the CO R1, 2, 3, and 4 lines can be judged very
similar to that of the R5 line. The profile of the R0 line is possibly
of the same type but blending is more severe.  The R5 profile is almost
a P Cygni profile: blue absorption accompanied by red emission.
However, the absorption component is not strongly blue shifted but 
has a velocity very similar to that of the absorption lines in
the other fundamental lines and to that of the 2-0 lines.

Absorption below the continuum is seen only in the 1-0 P and R branch
CO lines (Figure 3 -- 8). The observed interval includes the P1, P2, P3
and R0, R1, R2, R3, R4, and R5 lines with definite absorption below the
local continuum seen in all these lines except P1 and R0.  The strength
of the absorption at R5 suggests absorption below the local continuum
should be detectable to higher J lines of the R branch.  Due to the
isotopic shifts, the low J 1-0 lines of the isotopic variants are not
in the observed interval. The lowest member of the R branch in our
spectra is J'' = 9 for $^{13}$CO, J'' = 10 for C$^{18}$O, and J'' = 3 for
C$^{17}$O.  All the 1-0 lines regardless of isotopic species have a
stronger central valley than the vibrationally excited transitions.
The valley almost reaches the local continuum for the 1-0 $^{13}$CO
lines (note $^{13}$CO 1-0 R10 and R11 in Figure 7).

The central valley in the line profiles becomes asymmetric for the
stronger lines.  Self-absorption of the blue emission is obvious when
comparing the strength of the blue and red emission in profiles of the
1-0 CO lines (Figure 9).  The 1-0 CO line central absorptions are on
average 76\% broader on the blue side than the red side.  The weaker
$^{13}$CO 1-0 central absorptions are 20\% broader on the blue side.
The central absorption is systemically blue shifted relative to the
$\gamma$ velocity of the binary \citep{bakker_et_al_1998} for the very
lowest excitation lines.  The shift increases with decreasing J'', with
a shift of 1 km s$^{-1}$ for R5 and 3.5 km s$^{-1}$ for R2 (Figure
10).

The intensities of the emission lines of CO, $^{13}$CO, C$^{17}$O, and
C$^{18}$O  are remarkably similar and quite different from the
abundance ratios estimated from the 2-0 lines.  Peak intensities of the
following representative unblended lines illustrate this point:
\\
CO: 1-0 R2 28\%, 2-1 R11 27\%, 3-2 R11 11\% \\
$^{13}$CO: 1-0 R12 24\%, 2-1 R17 14\% \\
C$^{17}$O: 1-0 R11 9\%, 2-1 R17 14\% \\
C$^{18}$O: 1-0 R14 14\% \\

In contrast to the ratio CO/$^{13}$CO $\sim$ 10 from the 2-0 lines,
the CO/$^{13}$CO intensity ratio from 1-0 and 2-1 lines of similar
J is about 1.5.  Even more striking is the appearance of fundamental
lines of C$^{17}$O and C$^{18}$O with intensities about one-half
that of similar lines of CO.  Yet, CO/C$^{17}$O $>$ 100 from the
first-overtone lines.  The simplest interpretation of these contrasting
ratios is that emitting regions are optically thick in all the
observed fundamental lines. As is well known, the first-overtone
transitions are much weaker than the fundamental lines. Optically
thin emission in fundamental and first-overtone lines from a common
upper state  in the second vibrational level will differ by a factor
of about 100 in flux.  In the case of absorption from a common state
in the ground vibrational level, the absorption coefficient of the
1-0 line is similarly about  factor of 100 stronger than the 2-0
line.

The strengthening of the central absorption for the lowest energy
vibrational transition, the asymmetric absorption, and absorption
below the continuum require the presence of an absorbing gas cooler
than the emitting gas.  This absorbing gas has a velocity shifted
to the blue of the system barycentric velocity (-32.1 km s$^{-1}$)
by 1 to 3.5 km s$^{-1}$.  The emitting gas covers a $\sim$18 km
s$^{-1}$ range of velocity but is also shifted by $\sim$-1.5 km
s$^{-1}$ relative to the barycentric velocity.

\subsection{OH Fundamental}

The lowest excitation OH vibration-rotation 1-0 lines are in a
region of considerable telluric obscuration.  J''=4.5 is the lowest
OH level accessible under typical water vapor conditions (a few mm
of precipitable H$_2$O) at Gemini South.  However, a suitable order
sorting filter was not available for the J''=4.5 wavelength.  An
observation was made of the P branch line region for J''=5.5.  The
$^2\Pi$ OH ground state results in $\Lambda$-doubled rotational
levels, so each rotational line is divided into four components.
As a result, in spite of the large rotational line spacing for OH,
several OH lines can appear in a Phoenix spectrum taken with a
single grating setting.  The 3.0 $\mu$m 1-0 P$_{2f}$5.5 and
P$_{2e}$5.5 lines were detected in the spectrum of HR 4049 (Fig.
11).  This spectral region has considerable telluric absorption.
The removal of this absorption results in variable noise in the
ratioed spectrum.

The OH lines are, as are the CO lines, seen in emission.  The profiles
are similar to those of CO with double peaked profiles of observed FWZI
$\sim$26 km s$^{-1}$ centered at -35 km s$^{-1}$.  The two OH lines
observed are just 5\% above the continuum.  These were the only lines
that were detected in the 3.0 $\mu$m spectral region observed.

\subsection{H$_2$O Vibration-Rotation Lines} 

The asymmetric top molecule H$_2$O is known for the complexity,
apparent lack of rotational structure, and richness of its spectrum.
As a result H$_2$O lines are much more challenging to identify than
vibration-rotation lines of simple diatomic
molecules \citep{hinkle_barnes79}.  Emission lines are clearly present
in the 4.6 $\mu$m HR 4049 spectrum from three vibration-rotation
bands: $\nu_2$, $\nu_1-\nu_2$, and $\nu_3-\nu_2$.  With the above
caveats on the H$_2$O spectrum and based on the tentative identification
of four lines, the vibrationally excited band $2\nu_2-\nu_2$ possibly
also contributes to the 4.6 $\mu$m spectrum.  The strongest observed
H$_2$O transitions are from the $\nu_3-\nu_2$ band.  A number of
lines identified with this band have intensities $\sim$20\% above
continuum.  Typical H$_2$O lines are weaker than typical CO lines,
with many of the H$_2$O lines identified having intensity $<$10\%
above continuum.

Table 3 presents a list of the H$_2$O lines tentatively identified
in the HR 4049 spectrum.  In Figures 4 - 9 these lines are labeled
on the spectrum of HR 4049.  Many lines (e.g. Figure 9) are unblended
and clearly present.  However, a fairly large number are blended
with CO or other H$_2$O lines.  Due to the overlapping H$_2$O energy
levels, the line strengths of H$_2$O lines can vary significantly
between adjacent vibration-rotational transitions and, hence, the
contribution of a H$_2$O line to a blend is uncertain.

The band strength, $S {o \atop \nu}$, is a factor of five lower for the
$\nu_1-\nu_2$ band than the $\nu_3-\nu_2$ band.  However, the band
strength for the $\nu_2$ band is more than 10$^6$ higher than that of
either of these bands \citep{mcclatchey}.  The origin of the $\nu_2$
band is $\sim$1.5 $\mu$m red of the region observed.  While it would
be of interest to observe the lowest excitation $\nu_2$ lines, for
ground based observers the telluric $\nu_2$ lines are very strong and
prevent observations in 6 $\mu$m region. The $2\nu_2-\nu_2$ band has
similar band strength to the $\nu_1-\nu_2$ and $\nu_3-\nu_2$
combination bands but, unlike these bands which have origins in the
regions observed, $2\nu_2-\nu_2$ has an origin near that of
$\nu_2$.  This adds to our suspicion of the $2\nu_2-\nu_2$
identifications.

Like CO and OH lines the 4.6 $\mu$m H$_2$O lines have a double peaked
profile with emission peaks at -38.6 and -29.2 km s$^{-1}$ and
absorption at -33.6 km s$^{-1}$.  The observed FWZI of the weaker
H$_2$O lines is $\sim$25 km s$^{-1}$, perhaps slightly more narrow than
the CO lines.

In addition to the observed 4.6 $\mu$m transitions, H$_2$O also has
low excitation transitions in the 3.0 and 2.3 $\mu$m regions.  In
particular the $\nu_3$ band crosses the 3 $\mu$m region and has a
band strength similar to that of the $\nu_2$.  The $\nu_1$ band is
also present in the 3 $\mu$m region and, while weaker than $\nu_2$
or $\nu_3$, is a much stronger transition than the combination bands
seen at 4.6 $\mu$m.  Our 3.0 $\mu$m observation has an uneven
continuum perhaps as a result of weak emission features.  We undertook
a detailed search for H$_2$O lines but failed to identify any 3.0
$\mu$m H$_2$O lines.  Future searches of this region for H$_2$O
lines using higher signal-to-noise data and wider wavelength coverage
are justified.  On the other hand, our spectra in the 2.3 $\mu$m
region are of very high quality with broad wavelength coverage and
this region is clearly devoid of any contribution from H$_2$O.

\subsection{Line Profile Overview}

In summary of the above subsections, CO vibration-rotation fundamental
lines of four differently isotopically substituted species and
H$_2$O vibration-rotation lines populate the 4.6 $\mu$m region.
These lines all have double peaked emission profiles of total
(including both peaks) FWHM $\sim$ 19 km s$^{-1}$.  There is little
difference of intensity between lines of different isotopes.  The
lowest excitation lines, which are only seen in $^{12}$C$^{16}$O
in the wavelength range observed, have a central absorption as much
as 20\% below the local continuum. This central absorption overwhelms
the bluest of the double peaks in the emission profile but the
extreme bluest edge of the emission remains.  Examples of observed CO
fundamental and H$_2$O line
profiles are given in Fig. 9.  OH lines from the
fundamental vibration-rotation transition were seen in the 3 $\mu$m
region.  The OH lines are in emission with double peaked profiles
similar to those seen in the CO fundamental and H$_2$O lines.  The
CO vibration-rotation first-overtone transition appears in absorption
in the 2.3 $\mu$m region.  The absorption lines are nearly as broad
as the CO emission lines, FWHM $\sim$ 16 km s$^{-1}$, but have
simple Gaussian profiles and exhibit a range of line depths suggesting
the lines are optically thin.  $^{12}$CO dominates but weaker
$^{13}$CO is detectable.  The oxygen isotopes are not present.
All spectral lines in the 2-5 $\mu$m region have a small ($^>_\sim$~1 
km s$^{-1}$) shift blue of the systemic velocity.

\section{Modeling the Molecular Probes}

The fundamental spectrum presents a difficult analysis task.  The
CO is seen in both emission and, for the very lowest excitation
CO lines, absorption.  The small change in intensity
for emission lines over the full range of isotopes and over a large
range of excitation energy (rotational levels J"=0 to 30 and
vibrational transitions 1-0 to 3-2) clearly indicates that the
emission lines are very saturated.  A detailed investigation  of
these strongly saturated lines would require detailed radiative
transfer and disk modeling beyond the scope of this paper.

However,  analysis of the CO first overtone lines, which are seen
in absorption, is a much more tractable problem.  The observations
cover nearly all $^{12}$C$^{16}$O 2-0 R branch lines from J''=0 through
the highest detectable R branch line at J''=35.   The largest
interval of the 2-0 R branch not observed is R24 through R28.
The 2-0 P branch was observed from P1 through P8.  The 3-1 R
branch from J'' $\ge$ 4 also lies in the observed
region.  Equivalent widths of the first overtone CO line profiles
were measured from the fully processed, normalized spectra.  The
lines were measured both by summing the absorption area and second
by Gaussian fits to the line profiles.  Uncertainties were estimated
from the mean deviations from the Gaussian fits and by the formal
uncertainty in the fitted continuum level.

Equivalent width data was used to produce an excitation plot (Figure
12).  The log-linear increase of line strength with excitation level
demonstrates that the high-J v=2-0 lines (J$>$15) are optically thin
(i.e.  $\tau~<~$0.7). A least squares fit to the excitation plot of these
data requires a temperature of $\sim$550 K. However, the fit to these lines
underestimates the column density of the low-J lines.  To infer the
temperature and optical depth of the low-J lines, we extrapolated the
column density of the hot gas (inferred from the high-J lines) and
subtracted that from the measured column density. 

In order to correct for the effects of saturation, column densities
and level populations were determined from a curve of growth (COG)
analysis \citep[c.f.][]{spitzer_1978,brittain_2005}, which relates
the measured equivalent widths to column densities by taking into
account the effects of opacity on a Gaussian line profile.  The
derived column density for a measured equivalent width only depends
upon one parameter, the Doppler broadening of the line, $b$ =
$\sigma_{RMS}$/1.665, where $\sigma_{RMS}$ is the RMS linewidth.
To find the value of $b$, we apply two complementary methods:
comparison of the P and R branch lines and the linearization of the
excitation plot.  A key assumption is that the small scale line
broadening results entirely from thermal broadening.  The resolved
line profiles which are seen in the spectra indicate an additional
large scale broadening mechanism which will be discussed in \S5.1.

CO exhibits absorption lines in both P (J''=J'+1) and R (J''=J'-1)
branches, which have different oscillator strengths yet probe the same
energy levels, e.g., the P1 absorption line originates from the same
J=1 level as the R1 absorption line.  Any differences in the column
density derived from lines that share a common level must be due to
optical depth, which is related to $b$. The line width, $b$, can be
used to determine the optical depth and adjusted so that the derived
level populations from the two branches agree as closely as possible.

The (v=0, low-J) transitions are thermalized at densities as low as
n$_H$ $\sim$ 10$^{3-4}$ cm$^{-3}$, and at even lower densities due to
radiative trapping in the rotational lines with high opacity.
Therefore, the low-J lines are the ones most likely to exhibit a
thermal population distribution.  The line width that best linearizes a
plot of the level populations to a common temperature in an excitation
diagram is used.

Subject to the above constraints, the best fitting $b$ value in the
COG analysis for HR 4049 is 0.5$\pm$0.1 km s$^{-1}$.  The consistency
of all data to this common velocity dispersion is depicted in the
excitation plot of Figure 12.  With a measurement of $b$, equivalent
widths can be directly related to column density.  The column density
from fitting the 2-0 `high-J' lines is 4.6$\pm$0.3$\times$10$^{17}$
cm$^{-2}$ at a temperature of 530$\pm$20 K. The column density of
the `low-J' lines is 1.6$\pm$0.2$\times$10$^{18}$ cm$^{-2}$ at a temperature
of 40$\pm$10 K.  Uncertainty in the hot N($^{12}$CO) from the
overtone lines, estimated from the measurement of unsaturated lines,
is small and dominated by measurement errors in the equivalent
widths of the lines. The uncertainty in the cold gas is dominated
by the uncertainty in $b$.  Assuming that the 0.5 km s$^{-1}$ b
value for the cold gas applies to all the spectral lines, the opacity
of the most optically thick line is $\sim$1.5.  Increasing the b
value for the hot gas lowers the optical thickness of the higher
excitation lines.

The detection of weak 3-1 lines allows a check on vibrational LTE
in the gas. The column of CO in the v=1 state (from the v=3-1 lines)
is (5.7$\pm$1.3)$\times$10$^{15}$ and the temperature is 540$\pm$80K.
This is consistent with the temperature for the hot component of
the v=0 $^{12}$CO and $^{13}$CO branches (530$\pm$20 and 570$\pm$40
respectively). The combined 2-0 and 3-1 data give a rotational
temperature of 620$\pm$20K.  The vibrational temperature is
700$\pm$50K.  This is consistent with a slight overpopulation of
the v=1 state although the relative rotational populations are
consistent.  The vibrational temperature is more uncertain than the
other temperatures and evidence for non-LTE populations is weak.

Using the best fitting $b$ value, the column density can be determined
for other isotopic lines in the spectrum.  The corresponding column
density of $^{13}$CO is 2.3$\pm$0.3$\times$10$^{17}$ cm$^{-2}$ at a
temperature of 570$\pm$40 K.  Comparing lines of similar excitation,
$^{12}$C/$^{13}$C ratio is 6$^{+9}_{-4}$. First overtone C$^{17}$O
lines could not be detected.  The strongest lines, assuming a 550 K
excitation temperature, that are clear of both major telluric features
and blending CO lines are R5 and R8 (Fig.
2).  A firm upper limit on the equivalent width of these lines is
1.7 m\AA\ which translates to a column density
of 6$\times$10$^{14}$ cm$^{-2}$. At a temperature of 550 K, this
corresponds to a total column density of C$^{17}$O of less than
1$\times$10$^{16}$ cm$^{-2}$. Allowing for the temperature 
uncertainty a 3$\sigma$ limit for $^{16}$O/$^{17}$O is $>$200.

\section{Discussion}

The basic characteristics of the HR 4049 system are well understood.
At the heart of HR 4049 is a single-lined spectroscopic binary
\citep{bakker_et_al_1998}.  The visible early A/late B supergiant is a
low mass, perhaps white-dwarf mass, post-AGB star.   The unseen
companion is an M dwarf or white dwarf of lower mass than the
supergiant.  The infrared prominent feature of the HR 4049 system is
the circumbinary shell.  \citet{antoniucci_et_al_2005} review the
various geometries proposed for the circumbinary material.
Considerable evidence now points to a thick disk geometry.  Detailed
arguments are presented by \citet{dominik_et_al_2003}.

In the following discussion we adopt the \citet{dominik_et_al_2003}
disk model (Figure 1) with the following key points.  The disk is
optically thick with a height-to-radius ratio $\sim$1/3.  The dust on
the interior disk surface facing the star is approximately isothermal
at 1150 K.  The temperature of the dust wall implies a distance between
the star and dust of $\sim$10 AU.  The variability of HR 4049 and the
hydrostatic scale height suggests that the inclination of the disk is
$\sim$60$^\circ$ (i.e. the plane of the disk is tipped 30$^\circ$ from
the line of sight).  The optical depth of the dust, the height of the
disk, and the inclination result in only the far side of the disk being
observable (Figure 1).

\subsection{Circumbinary Flow}

Previous observations of the CO first overtone are reported by
\citet{lambert_et_al_1988}.  Based on an excitation temperature of
300 $\pm$ 100 K and a non-stellar velocity \citet{lambert_et_al_1988}
conclude the CO is circumstellar.  The much higher resolution and
S/N data analyzed above refine the excitation temperature to 520
$\pm$ 20 K for the higher excitation lines and 40$\pm$10 K for
the low excitation lines.  The velocities reported here and those
reported by \citet{lambert_et_al_1988} show no change over nearly
20 years, as expected for lines of circumbinary origin.  Although
of the current data is of higher precision, both data sets are
consistent with an outflow velocity of $\sim$1 km s$^{-1}$.  The
column densities reveal that about four times more cool gas,
$\sim$2$\times$10$^{18}$ cm$^{-2}$, is present than hot gas,
$\sim$5$\times$10$^{17}$ cm$^{-2}$.

The observations demonstrate that the gas is in rotational LTE and
near or in vibrational LTE.  For vibrational equilibrium the critical
density is n$_H~\sim$ 10$^{10}$ cm$^{-3}$ \citep{najita_et_al_1996}.
Taking this density and a CO column density of 2$\times$10$^{18}$
cm$^{-2}$, the thickness of the CO absorption line forming region
is $\sim$4$\times$10$^{11}$ cm.  So the gas is restricted to a
zone radially $\sim$6 R$_\odot$ from the disk inner dust wall.
The CO appears to depart slightly from vibrational LTE, so the
density is likely slightly lower than the critical density.  In any
case, the thickness of the gas layer is certainly thin compared to
the 2150 R$_\odot$ spacing between the binary and dust wall.

If the gas layer is located radially just on the star side of the
dust wall, adopting the \citet{dominik_et_al_2003} geometry permits
the total gas mass to be calculated.  Taking the radius to be 10
AU and the height of the disk to be 1/3 the radius, the surface
area of the cylindrical wall follows.  The column density then gives
the total number of CO molecules.  Since the gas is oxygen rich,
the number of CO molecules is limited by the carbon abundance.
Taking [C/H] for HR 4049 from \citet{waelkens_et_al_96} and
the solar carbon abundance of \citet{grevesse_et_al_1991}, the mass of
the gas disk is 6 $\times$10$^{26}$ gm, i.e. $\sim$ 0.1 M$_\oplus$.
A total disk mass of a $ > \atop \sim$ 33 M$_\oplus$ was suggested
by \citet{dominik_et_al_2003}, so the mass estimates are in accord
with a thin gas zone at the edge of a more massive dust disk.

The CO first overtone lines
are symmetric and $\sim$16 km s$^{-1}$ across.  In contrast, a line
width 20 times smaller, $\sim$0.5 km s$^{-1}$, is required to model
the curve of growth.  A 0.5 km s$^{-1}$ width is consistent with
thermal broadening.  We suggest that the broadening to 16
km s$^{-1}$ is due to a systemic flow of gas in the circumstellar
shell.  As noted above, our view of the dust disk continuum is
limited to the side opposite from the star (Figure 1), so many kinds
of axisymmetric flows could result in the observed line broadening.
We consider the line shapes to constrain further our
understanding of the flow.

All the unblended 4.6 $\mu$m emission line profiles, including those
for weaker lines, are double peaked.  Since $all$ the lines are
double peaked, we discount self-absorption as the principal cause
for this line shape.  Double peaked lines suggest an origin in a
rotating ring or disk.  The observation of CO fundamental band
absorption demands a P-Cygni type geometry, i.e., an emitting area
extended relative to the continuum forming area.  While emission
is very dominant in the 4.6 $\mu$m HR 4049 spectrum, there is a
hint of underlying and offset absorption in all the lines with the
emission line profiles having a lower peak on the blue side than
on the red side.  The spectrum is dominated by saturated lines
10-20\% above the continuum.  However, some lines are stronger and
we assume these stronger lines result as optical thick transitions
cover larger areas.  At 3 $\mu$m radiation from a 550 K blackbody
is about half that at 4.6 $\mu$m and, indeed, the 3 $\mu$m OH
$\Delta$v=1 lines are $\sim$5\% above continuum.  Ultimately
saturation combined with the physical extent of the line forming
region demands a limited range of emission line strengths.  The
strongest lines in the 4.6 $\mu$m CO spectrum of HR 4049 are
$\sim$40\% above the dust continuum.

The simple model of a rotating disk can be applied to the existing
disk model (Figure 1) and tested by producing model profiles of the
lines.  The CO first overtone lines suggest that most of the CO
occurs in a relatively thin layer.  An absorption line was modeled
by assuming a continuum source of 10 AU radius.  The layer of
absorbing gas was divided into zones of 0.1 AU along the circumference.  The line
RMS was assumed to be 0.5 km s$^{-1}$ and the gas was assumed to
be rotating in a Keplerian orbit. The resulting profile is a double
peaked absorption line. This profile was then convolved with a
Gaussian instrumental profile with 6 km s$^{-1}$ FWHM.  The
resulting synthetic profile, which is an excellent match to
observations, is shown in Figure 13.

Is this model also consistent with the 4.6 $\mu$m emission line shapes?
To investigate this question five assumptions were made: (1) The
emissivity of the gas is constant over the entire region modeled.
(2) The gas is in Keplerian orbits.  (3) Line broadening is limited
to the thermal b value, 0.5 km s$^{-1}$ discussed above and the
broadening from the Keplerian motion. (4) Absorption is insignificant.
(5) The gas originates at 10 AU and extends to larger radii.  Since
the dust disk is opaque this extension is along the top of the disk
(Figure 1).  An extension to larger radii was included since there 
is no requirement for a background continuum source for the emission 
line spectrum. To fit the profiles with this model we found a maximum
radius of 14 AU.  As for the overtone model, the disk was divided
into zones, the profile from each zone shifted and weighted by the
viewing aspect, and then summed into velocity bins of 0.1 km s$^{-1}$.
The resultant double-peaked profile of the emission line is seen
in Figure 14.  Convolution with a Gaussian instrumental profile of
FWHM = 6 km s$^{-1}$ produced a good match to a typical emission
line.

While consistency between the modeled and observed profiles is
satisfying, the fundamental transitions clearly require much more
refined modeling to address a number of details.  For instance,
there is a large difference between the 550 K CO and 1150 K dust
temperatures.  If, as is commonly assumed \citep[see
e.g.][]{glassgold_et_al_2004}, the gas and dust temperatures are
in equilibrium within the disk then the 1150 K dust temperature
applies only to a surface layer.  Radiative cooling
from CO fundamental emission \citep{ayres_wiedemann_1989} is largely
disabled by the large optical depth of the CO lines
\citep{glassgold_et_al_2004}.  It is plausible that the gas undergoes
heating on exiting the disk.

In an isothermal model optically thick CO self-absorption occurs
for the fundamental transitions; the opacity in the low J 1-0 is
$\sim$400.  The fundamental lines are seen in emission because the
the 550 K temperature of the CO makes optically thick CO lines
brighter at 4.6 $\mu$m than the 1150 K continuum.  At the resolution
of the observations, 6 km s$^{-1}$, narrow self-absorption lines
of 0.5 km s$^{-1}$ width are largely smeared out.  Additionally,
the gas is certainly not isothermal.  If, for instance, the gas is
heated as it leaves the disk, the temperature profile could increase
toward the observer.  Depending on the details of the spatial
filling, optical depth, and temperature profile absorption is not
a requirement.

Two temperatures were measured in the CO first overtone, $\sim$40
and $\sim$550 K but no velocity differences were measured between
the 40 K and 550 K regions, suggesting that these temperature regions
are physically close together.  Both the 40 K and 550 K CO are seen
in absorption against the 1150 K continuum.  H$_2$O, on the other hand,
is seen only
in emission.  Emission lines are not spatially limited to the 1150
K continuum forming region.  If the absence of H$_2$O absorption
results from H$_2$O existing only in the disk edge region and not
in the disk mid-plane, the gas is differentiated vertically as well
as horizontally relative to the plane of the disk.

The
measured decrease in the column density in between the v=0 and v=1
levels implies that there is ample population to produce the observed
optically thick 2-1 lines.  However, the observation of optically
thick 3-2 emission suggests that the v=3 level is populated
above that expected from LTE.  Overtone transitions
higher than 3-1 are outside of the region observered.  It would be
of interest to search for the strongest lines in the 4-2 band.

The emission profile is shifted relative to the center-of-mass
velocity by $\sim$1.5 km s$^{-1}$, suggesting an outflow.  If the
depression of the blue wing in the emission profiles is due to
absorption in front of the dust disk, the absorption line profile
is formed in a region with less outflow than the extended emission
line forming region.  Outflow was also noted for the first overtone
CO lines.  The outflow increases for the very lowest excitation
lines, suggesting that the gas cools in the inner-disk region and
is accelerated as it flows out.  For the lowest excitation CO
fundamental lines the cold outflow is seen in absorption with
the outflow velocity increasing (Figure 10) as excitation energy decreases. 
\citet{dominik_et_al_2003} suggested that along the edges of the
disk an outward flow results from radiation pressure erosion.
Alternatively, a disk pressure gradient can result in an outward
flow \citep{takeuchi_lin_2002} without a need for small grains.
Indeed, the outflow could be driven initially by either gas or dust
since momentum is transferred between the gas and dust through
collisions \citep{netzer_elitzur_1993}.  

\subsection{Comparison with Optical Spectra}

A detailed analysis of time series C I, Na I D (D$_1$ and D$_2$),
and H $\alpha$ spectra of HR 4049 is presented by
\citet{bakker_et_al_1998}.  The Na D lines contain a number of
absorption components as well as weak emission.  \citet{bakker_et_al_1998}
identify two Na D absorption components with the circumstellar
environment of the binary system.  These are labeled as `A$_1$' and
`A$_2$' \citep[see Table 2 and Figure 4 of][]{bakker_et_al_1998}.
A$_1$ has a velocity of $\sim$ -5.0 km s$^{-1}$ (mean of D$_1$ and
D$_2$) relative to the systemic velocity.  A$_2$ has a velocity of
-0.8 km s$^{-1}$ again relative to the systemic velocity.  A$_2$
is stronger than A$_1$ by about 50\% and has a slightly greater
FWHM.

The continuum in the infrared is dominated by the dust continuum.
However, in the optical the continuum is entirely from the stellar
photosphere.  Thus the optical absorption is formed along a pencil
beam originating near the center of the circumbinary disk.  
The A$_1$ velocity has similar velocity to the outflow seen in the lowest
excitation 1-0 CO lines.  This outflow is a cold wind perhaps leaving
the system.  The A$_2$ outflow is close to the outflow velocity
seen in the CO first overtone as well as the slightly higher
excitation 1-0 lines (Figure 10).  This flow is an outward flow of
warmer gas perhaps associated with circulation in the disk.

Given that the star is the continuum source of the Na absorption
and the rear inner walls of the disk are the continuum source for
the CO absorption, perfect agreement is not expected in either line
of sight velocity or in FWHM.  The overtone CO has a much larger
FWHM than the Na D, as expected given the larger range of velocities
sampled by the CO along the lines of slight to the CO continuum
forming area (Fig. 13).

Na I also has an emission component \citep[`A$_3$'
in][]{bakker_et_al_1998}.  This is perhaps due to fluorescent
emission from the gas interior to the disk.  The line profile is
disrupted by the Na D absorption components but the FWMH of the
emission, $\sim$21 km s$^{-1}$, is comparable to that of the CO
emission.

The H$\alpha$ line profile is complex.  \citet{bakker_et_al_1998}
identified two components, `C$_{max}$' and `R$_{min}$' which are
stationary and presumably are associated with the circumbinary
environment.  Both are seen in absorption.  C$_{max}$ has a large
outward velocity, -21.3 km s$^{-1}$.  The velocity of R$_{min}$ is
much less, -7.5 km s$^{-1}$.   The energetics of the H$\alpha$ line
are very different from those of the cold gas lines discussed in
this paper.  H$\alpha$ also has an absorption feature `B$_{min}$'
which possibly varies in anti-phase with the primary.  Detailed
understanding of the excited gas sampled by H$\alpha$ requires
modeling beyond the current discussion.

\subsection{Properties of the Binary Members}

The conceptual picture of a thin gas layer co-rotating just in front
of the dust wall suggests the observed velocities result from
Keplerian rotation.  A FWZI of 18 km s$^{-1}$ implies a rotational
velocity of 9 km s$^{-1}$.  Assuming Keplerian rotation and a 10
AU disk radius, the total binary mass required is 0.9 M$_\odot$.
This is in agreement with the total binary mass suggested by
\citet{bakker_et_al_1998}.  Bakker's mass was based on the mass
function from the spectroscopic orbit and the assumption that the
A supergiant had a typical white dwarf mass.

The mass function from \citet{bakker_et_al_1998}, the total binary
mass, and the orbital inclination allows a solution for the individual
masses in the binary.  We make the assumption, discussed in \S5.6,
that the binary orbit is co-planer with the disk.  The definition
of the mass function then yields the masses for the individual
stars.  The A supergiant has a very low mass of 0.58
M$_\odot$ confirming the post-AGB state of
this star.  This mass, nearly equal to that of a typical white dwarf
\citep{bergeron_et_al_1992}, implies that the mass-loss process for
this star has terminated.  The companion mass is 0.34 M$_\odot$.
This mass does not resolve the status of the companion.  While a
mass of 0.34 M$_\odot$ is low for a white dwarf and strongly suggests
an M-dwarf, it is possible that the companion mass has been altered
by evolution (\S5.6).

\subsection{Winnowing}

\citet{lambert_et_al_1988} report quantitative abundances for HR
4049 revealing an extremely metal-poor star with [Fe/H] $< \atop
\sim$ -3 but near-solar C, N, and O: [C/H] = -0.2, [N/H]=0.0,
[O/H]=-0.5.  \citet{lambert_et_al_1988} argue that the ultra-low
iron abundances found in a post-AGB star cannot be primordial since
there are no known progenitor AGB stars with similar abundances.
\citet{venn_lambert_1990} and \citet{bond_1991} find similar abundance
patterns to those in HR 4049 in the young main-sequence $\lambda$
Boo stars and gas in the interstellar medium (ISM).  In all three
cases the abundance pattern is deficient in refractory (high
condensation temperature) elements but nearly solar in volatile
(low condensation temperature) elements.  This abundance pattern
is explained in the ISM by the locking up of refractory elements
in grains.

Five extremely iron-deficient post-AGB stars are known in the HR
4049 class \citep{van_winckel_et_al_95}.  \citet{lambert_et_al_1988}
and \citet{mathis_lamers_1992} have noted that all are A stars with
no surface convection.  A likely scenario is that the observed
abundances result from peculiar abundances in little more than the
observed photospheric layer.  The very low refractory abundance in
the HR 4049 stars results in a much lower opacity in the photospheric
material than from a normal composition making this region additionally
stable against convection \citep{mathis_lamers_1992}.  Assuming a
47 R$_\odot$ radius for HR 4049 \citep{bakker_et_al_1998} and
referring to a 7500 K T$_{eff}$, log g = 1.0 model atmosphere
\citep{kurucz_1979}, the photosphere of HR 4049 above optical depth
unity contains a few percent of an Earth mass of volatile material.

\citet{mathis_lamers_1992} postulated that the HR 4049 abundance
pattern results from the separation of mass-loss gas and dust by
differential forces on the gas and dust in a circumstellar shell.
\citet{waters_et_al_1992} further suggested that a circumbinary
disk played a critical role.  Winnowing occurs as gas is accreted
to the stellar surface while dust remains in the circumstellar shell
or is ejected.  The $\lambda$ Boo stars, which are also A stars
without surface convection, have similar surface abundances due to
winnowing of gas from dust in a pre-main sequence disk
\citep{venn_lambert_1990}.  \citet{mathis_lamers_1992} found the
removal of refractory elements from a solar abundance gas to be a
very inefficient winnowing process.  They suggested that an efficient
winnowing process is one that creates a gas with a low refractory
abundance.

Models of disks, created mainly to explore pre-main sequence
evolution, provide a rich view of the basic disk physics.  While
the detailed physics of the winnowing are complex, these models,
combined with the current observations, reveal the basics of the
winnowing process.  In optically thin circumstellar regions, the
radiation pressure to gravity ratio for A stars drags grains with
larger than 4 microns inward while expelling grains smaller than 4
microns \citep{takeuchi_artymowicz_2001}.  \citet{takeuchi_lin_2002}
extend this to disk models, showing that large particles accumulate
in the inner part of the disk.  These models also apply to optically
thick disks where the majority of the dust in the disk is not exposed
to stellar radiation \citep{takeuchi_lin_2003}.  In this case,
interaction with the stellar radiation field at the inner disk edge
drives flows in the disk with the dust-to-gas ratio increasing at
the inner disk edge.

\citet{dominik_et_al_2003} conclude that the grain size distribution
in the HR 4049 disk is currently indeterminate.  However, they note that in
the case where the inner disk consists of small grains, these grains
will be driven outward by radiation pressure exposing a dust free
region of gas.  This gas layer will be driven inwards by either the
gas pressure gradient or sub-Keplerian rotation.  In pre-main
sequence disks it has been shown that the gas interior to the dust
suffers turbulent viscosity and accretes onto the central star.
The  viscous time scale in pre-main sequence circumstellar disks
is typically estimated at $\sim$ 10$^6$ years
\citep{takeuchi_lin_2003,hartmann_et_al_1998}.

The observations reported here of a sheet of gas at the inner disk
edge support the model where separation of volatiles from grains
occurs near the inner dust disk surface.  The temperature of the
gas released from the grains is far to low for evaporation of
refractory elements to take place.  The total gas mass interior to
the HR 4049 dust disk is $\sim$0.1 M$_\oplus$.  The gas
required in the stellar photosphere to alter the observed stellar
abundances is one tenth this.  A naive interpretation is
that the surface material required to match the observed abundances
could be accreted in $< \atop \sim$10$^5$ years.  For
post-AGB evolution this may be too long.  An alternative is
that the winnowing process currently observed is the termination
of a very rapid clearing of the inner disk region that result in
sudden accretion of the gas now present in the stellar photosphere.
The observed CO, H$_2$O, OH is near the disk.  The flow
to the stellar surface is presumed much more tenuous and is not
observed.

\citet{takeuchi_lin_2002} found that higher than a few disk scale
heights from the disk midplane the gas rotates faster than the
particles due to an inward pressure gradient.  This drag causes
particles to move outward in the radial direction.
\citet{takeuchi_lin_2003} speculate that in an optically thick disk,
particles in the irradiated surface layer move outward, while beneath
the surface layer, particles move inward.  
The outward flow seen in CO plus the driving entrained particles
could 
rejoin the cool outer portions of the disk.  In this case, the inward
interior disk flow would move this material to the inner disk surface.  The
winnowing process could then be a distillation process resulting
in a disk with increasingly refractory grains.

\subsection{Isotopic Abundances}

The oxygen isotopic ratios reported by \citet{cami_yamamura_2001}
set HR 4049 apart as having by a factor $>$10 the smallest ratios
of $^{16}$O/$^{17}$O and $^{16}$O/$^{18}$O known at that time.  Our
analysis of the optically thin CO first overtone transition does
not support these results.  There are no detectable C$^{17}$O first
overtone lines giving a 3$\sigma$ limit of $^{16}$O/$^{17}$O $>$
200.  On the other hand, the fundamental spectrum of CO in HR 4049
consists of optically thick emission lines.  Four isotopic variants
($^{12}$C$^{16}$O, $^{13}$C$^{16}$O, $^{12}$C$^{17}$O, and
$^{12}$C$^{18}$O) can be seen in the fundamental spectrum with lines
of similar intensity.  The CO$_2$ bands measured by
\citet{cami_yamamura_2001} were observed at low resolution by ISO
and are in the 13 - 17 $\mu$m region of the infrared.  These bands
appear in emission.  We contend that the oxygen isotope ratios
appear small in these CO$_2$ bands because, as for the CO fundamental,
the emission lines are highly saturated.  \citet{cami_yamamura_2001}
warn that their isotopic ratios are in the optically thin limit.

The carbon and oxygen isotopic ratios appear typical for an AGB
star \citep{lambert_1988}.  While the oxygen isotopic ratio in the
circumstellar environment of HR 4049 is not abnormally low, there
are stars that do have extreme oxygen isotope values.  Some hydrogen
deficient carbon stars have $^{16}$O/$^{18}$O
considerably less than 1 \citep{clayton_et_al_2005, clayton_et_al_2007}.
\citet{clayton_et_al_2007} suggest that the extreme overabundance
of $^{18}$O observed in these objects is the result of He-burning
in white dwarf mergers.  Meteoritic samples have been found with
small $^{16}$O/$^{18}$O ratios.  These could result from processes
in the pre-solar nebula or pollution from a stellar source.  While
the rarity of stellar sources with small oxygen isotope ratios
suggests a stellar source is unlikely, the origin of exotic oxygen
isotopic ratios detected in early solar system samples remains
uncertain \citep{aleon_et_al_2005}.

\subsection{Binary Evolution}

ISO data described by \citet{dominik_et_al_2003} contain features
from oxygen-rich molecules implying that the disk is oxygen-rich.
The lack of mid-infrared silicate features associated with oxygen-rich
grains is attributed by \citet{dominik_et_al_2003} to high optical
depth in the dust disk.  Our observations reveal a 2.3 -- 4.6 $\mu$m
spectrum resulting from a mix of gas phase molecules, CO, OH, and
H$_2$O, typical for an oxygen-rich environment.  If, as seems probable, 
the gas consists of volatiles evaporated from grains then
the disk environment is oxygen rich.

In contrast, as reviewed in \S1, carbon-rich circumstellar grains
have been observed.  An explanation is that these grains are exterior
to the disk.  Carbon-rich chemistry is the result of evolution in
the AGB phase where CNO material processed in the stellar interior
is mixed to the surface.  For AGB stars of mass $ > \atop \sim$ 2,
the surface chemistry of the AGB star is converted to carbon-rich
by the third dredge up.  Rapid AGB mass loss then produces a
carbon-rich circumstellar shell.  In the case of HR 4049 the fossil
carbon-rich shell of AGB mass loss is still observable
although HR 4049 is now a post-AGB object.

Why is the disk oxygen-rich?  \citet{bakker_et_al_1996} noted that
the current binary separation is less than the radius required for
the AGB phase of the current post-AGB star.  Hence, prior to the
post-AGB stage the HR 4049 system underwent common envelope evolution.
Prior to the common envelope phase the system passed through a
pre-AGB contact binary phase with the more massive star transferring
mass onto the less massive member.  An AGB star does not contract
due to mass loss, so the AGB star continued to expand enveloping
the dwarf companion.  Common envelope systems rapidly eject mass
from both members \citep{taam_sandquist_2000}.  Most carbon stars
have C/O near unity \citep{lambert_et_al_1986}.  Mixing or mass
transfer during the common-envelope stage converted the carbon-rich
envelope of the AGB star back to an oxygen-rich envelope.  Mass
lost during the common-envelope phase formed the current circumbinary
disk.  In such a scenario a co-rotating circumbinary disk is formed
surrounding the binary \citep{rasio_livio_1996}.

This process is apparently not unusual.  \citet{de_ruyter_et_al_06}
find that circumbinary disks are a common feature of post-AGB stars.
The current A-supergiant M-dwarf HR 4049 binary is rapidly evolving
to a white-dwarf M-dwarf binary system.  White-dwarf M-dwarf binary
systems with a co-binary disk resulting from common-envelope evolution
also appear to be common \citep{howell_et_al_2006}.

\section{Conclusions}

The 2 to 5 $\mu$m spectrum of HR 4049 is formed in a circumbinary
disk and wind.  The optically thin 2.3 $\mu$m CO lines appear in
absorption against the dust continuum, allowing the determination
of the mass of gas.  The gas forms a thin layer, of radial thickness
$\sim$ 6 R$_\odot$, lining the dust disk.  This gas is 
composed of the volatiles separated in the disk from grains.  The 
4.6 $\mu$m emission
spectrum requires a region of line formation extended beyond the
continuum forming region.  The circumbinary gas is rotating with a
Keplerian velocity of $\sim$ 9 km s$^{-1}$.  Combined with the
circumbinary disk radius and inclination derived from photometry,
Keplerian rotation allows the determination of the masses of the
individual binary stars.  The very low mass of the A supergiant,
0.58 M$_\odot$, confirms the post-AGB nature of this object.  Gas
is also flowing out of the system, perhaps as a result of a disk
pressure gradient, at $\sim$1 km s$^{-1}$.

Our observations show that the HR 4049 circumbinary disk has typical
AGB abundances for the carbon and oxygen isotopes;  $^{12}$C/$^{13}$C
= 6$^{+9}_{-4}$ and $^{16}$O/$^{17}$O $>$200.  Exotic mechanisms,
as proposed e.g.~by \citet{lugaro_et_al_2004}, for production of
the $^{16}$O/$^{17}$O are not required.  The widely quoted value
of $^{16}$O/$^{17}$O $\sim$ 8 reported by \citet{cami_yamamura_2001}
results from the naive interpretation that the infrared emission
lines are optically thin.  \citet{cami_yamamura_2001} warn that
their values were in the optically thin limit.  The extreme saturation
of lines in the 4.6 $\mu$m spectrum of HR 4049 results in nearly
equal apparent strengths for isotopic variants of molecular species
with abundances differing by factors of 10$^3$.

The peculiar surface abundances of HR 4049 are likely the result
of winnowing driven by the evaporation of volatiles in the disk and
the viscous accretion of this gas onto the star.  Detailed modeling
of the process will be required to determine if there is adequate
time for the current outgassing of the disk to fully explain the
surfaces abundances of the A supergiant or if a sudden, post-common
envelope clearing of the inner disk is required.  The existence of
the $\lambda$ Boo stars shows that the winnowing process applies
to pre-main sequence as well as post-AGB systems.  The wider
significance of the winnowing process may well be in systems where
the convective nature of the stellar photosphere cancels any impact
on stellar abundances.  However, circumstellar grains in these
systems are undergoing processes separating 
volatile and refractory elements.  This winnowing could have general
application to the chemical evolution of grains in pre-main sequence
disks.

Carbon-rich circumstellar material implies that the post-AGB star
was a carbon-rich star on the AGB.  The current oxygen-rich
circumstellar disk likely evolved from common envelope mixing.  HR
4049 is one of five known post-AGB stars with similar photospheric
abundances.  Of the other four at least one, the red-rectangle
nebula/binary HD 44179,  has a similar oxygen-rich circumbinary
disk in a carbon-rich circumstellar shell \citep{waters_et_al_1998}.
In future papers of this series we will explore the infrared spectra
of the other members of the HR 4049 class of objects.

\acknowledgements

This paper is based in part on observations obtained at the Gemini
Observatory, which is operated by the Association of Universities
for Research in Astronomy, Inc., under a cooperative agreement with
the NSF on behalf of the Gemini partnership: the National Science
Foundation (United States), the Particle Physics and Astronomy
Research Council (United Kingdom), the National Research Council
(Canada), CONICYT (Chile), the Australian Research Council (Australia),
CNPq (Brazil), and CONICRT (Argentina).  The observations were
obtained with the Phoenix infrared spectrograph, which was developed
and is operated by the National Optical Astronomy Observatory.  The
spectra were obtained as part of programs GS-2002A-DD-1, GS-2002B-DD-1,
GS-2003A-DD-1, GS-2004A-DD-1, and GS-2005B-DD-1.  We thank Drs.
Claudia Winge and Bernadette Rodgers and the Gemini South staff for
their assistance at the telescope.  We thank Dr. Richard Joyce for 
useful discussions.  We thank the anonymous referee
for a very detailed critical reading of the draft.  S.D.B.  acknowledges
that work was performed under contract with the Jet Propulsion
Laboratory (JPL) funded by NASA through the Michelson Fellowship
Program. JPL is managed for NASA by the California Institute of
Technology.


\clearpage \begin{deluxetable}{lccc} \tablenum{1} \tablewidth{0pt}
\tablecaption{Log of Observations} \tablehead{ \colhead{Date} &
\colhead{Wavelength}  & \colhead{Frequency}  & \colhead{S/N} \\
\colhead{} & \colhead{($\mu$m)}  & \colhead{(cm$^{-1}$)} &  \colhead{}
} \startdata 2002 Feb 13 & 2.3120 & 4324 & 290 \\ 2002 Feb 13 &
2.3233 & 4303 & 220 \\ 2002 Feb 13 & 2.3309 & 4289 & 280 \\ 2002
Feb 13 & 2.3421 & 4268 & 260 \\ 2002 Feb 13 & 2.3617 & 4233 & 400
\\ 2002 Dec 11 & 2.3405 & 4272 &  80 \\ 2002 Dec 12 & 2.3126 & 4323
&  90 \\ 2002 Dec 12 & 4.6434 & 2153 & 200 \\ 2002 Dec 12 & 4.6629
& 2144 & 190 \\ 2002 Dec 14 & 2.2980 & 4350 & 350 \\ 2002 Dec 14 &
2.9977 & 3335 & 130 \\ 2002 Dec 14 & 4.6219 & 2163 & 150 \\ 2002
Dec 14 & 4.6825 & 2135 & 170 \\ 2003 Feb 16 & 2.3416 & 4269 & 210
\\ 2004 Apr 3  & 4.8874 & 2045 & 220 \\ 2005 Dec 10 & 2.3634 & 4230
& 300 \\ 2005 Dec 10 & 2.3523 & 4250 & 350 \enddata \end{deluxetable}

\begin{deluxetable}{llclclclc} \tablenum{2} \tablewidth{0pt}
\tablecaption{CO $\Delta$v=1 Line List} \tablehead{\colhead{Species}
&
	    \colhead{Line} & \colhead{i$_c$$^{1}$} & \colhead{Line}
	    & \colhead{i$_c$} & \colhead{Line} & \colhead{i$_c$} &
	    \colhead{Line} & \colhead{i$_c$}
	  }
\startdata $^{12}$C$^{16}$O & 1-0 P3 & 1.36: & 1-0 P2 & 1.25 &
1-0 P1 & 1.27 & 1-0 R0 & 1.28 \\
	 &
1-0 R1 & 1.26  & 1-0 R2 & 1.25 & 1-0 R3 & 1.17 & 1-0 R4 &
1.36 \\
	 &
1-0 R5 & 1.38  &       &       &      &      &       &      \\
	 &
2-1 P18 &  1.36 & 2-1 R3 & \nodata  &    2-1 R4 & 1.27      &
2-1 R6 & \nodata  \\
	 &
2-1 R7 & 1.37 & 2-1 R8 & 1.31 & 2-1 R8 & 1.31 & 2-1 R9 &
1.26 \\
	 &
2-1 R10 & \nodata & 2-1 R11 & 1.27 & 2-1 R12 & 1.27 & 2-1
R13 & 1.22 \\
	 &
3-2 P11 & 1.13  & 3-2 R10 & \nodata & 3-2 R11 & 1.08 & 3-2
R12 & \nodata \\
	 &
3-2 R13 & \nodata  & 3-2 R14 & 1.08   & 3-2 R15 & \nodata  &
3-2 R16 & \nodata \\
	 &
3-2 R18 & \nodata & 3-2 R21 & 1.09 &   &   &  &  \\
	 &
5-4 P5 & 1.03  &   &   &   &   &   &   \\
	 & &       &   &   &   &   &   &   \\
$^{13}$C$^{16}$O & 1-0 R9 & \nodata & 1-0 R10 & 1.23 & 1-0 R11
& 1.20 & 1-0 R12 & 1.23 \\
	 &
1-0 R13 & 1.22 & 1-0 R14 & \nodata & 1-0 R15 & 1.27 & 1-0
R16 & 1.20 \\
	 &
1-0 R17 & 1.20 & 1-0 R18 & 1.21 & 1-0 R19 & 1.17 & 1-0 R20
& \nodata \\
	 &
2-1 P7 & 1.28  & 2-1 P6 & 1.25 & 2-1 R17 & 1.12 & 2-1 R18
& \nodata \\
	 &
2-1 R19 & 1.16 & 2-1 R20 & \nodata & 2-1 R21 & \nodata & 2-1
R22 & 1.13 \\
	 &
2-1 R23 & \nodata & 2-1 R24 & \nodata & 2-1 R25 & 1.13 & 2-1
R26 & \nodata \\
	 &
2-1 R27 & 1.09 & 2-1 R28 & \nodata & 2-1 R29 & \nodata & 2-1
R30 & 1.08 \\
	 &
	  &       &   &   &   &   &   &   \\
$^{12}$C$^{17}$O & 1-0 P17 & 1.14 & 1-0 R3 & \nodata & 1-0 R4
&  1.11 & 1-0 R5 &  1.15 \\
	 &
1-0 R6 & \nodata & 1-0 R8 & \nodata & 1-0 R9 & 1.12 & 1-0
R10 & \nodata \\
	 &
1-0 R11 & 1.08 & 1-0 R12 & \nodata & 1-0 R13 & \nodata &   &
\\
	  &
2-1 R17 & \nodata & 2-1 R19 & \nodata &        &         &   &
\\
	 &
	  &       &   &   &   &   &   &   \\
$^{12}$C$^{18}$O & 1-0 P13 & 1.26 & 1-0 P12 & 1.24 & 1-0 R10
& 1.12 & 1-0 R11 & 1.14 \\
	 &
1-0 R12 & \nodata & 1-0 R13 & \nodata & 1-0 R14 & 1.12 & 1-0
R15 & 1.12 \\
	 &
1-0 R16 & 1.18 & 1-0 R17 & 1.13 & 1-0 R18 & 1.11 & 1-0 R19
& \nodata \\
	 &
1-0 R21 & 1.13 & 1-0 R22 & 1.12 &  &  &  &  \\
	  &
2-1 R22 & 1.03 &    &   &   &   &   &   \\ \enddata

\tablecomments{Central intensities at the maximum emission strength.
Continuum = 1.0.} \end{deluxetable}

\begin{deluxetable}{ccccccc} \tablenum{3} \tablewidth{0pt}
\tablecaption{H$_2$O Line List} \tablehead{ \colhead{Vibrational}
&
	    \colhead{Rotational}  & \colhead{i$_c$} & \colhead{Rotational}
	    & \colhead{i$_c$} & \colhead{Rotational}  & \colhead{i$_c$}
	    \\ \colhead{Transition}     & \colhead{Transition}     &
	    \colhead{} & \colhead{Transition}     & \colhead{} &
	    \colhead{Transition}     & \colhead{}
	  }
\startdata

(100)-(010) & [2,2,0]-[1,1,1]  & 1.07 & [2,2,1]-[1,1,0] & 1.19: &
[3,1,3]-[2,0,2]  & \nodata \\
	    &
[3,2,1]-[4,1,4]  & 1.13  & [3,2,2]-[2,1,1]  & \nodata & [4,0,4]-[3,1,3]
& \nodata \\
	    &
[4,1,4]-[3,0,3]  & 1.16 & [4,2,3]-[4,1,4]  & 1.11 & [5,0,5]-[4,1,4]
& 1.13  \\
	    &
[5,1,5]-[4,0,4]  & 1.10 & [5,2,4]-[5,1,5]  & \nodata & [5,4,2]-[5,3,3]
& 1.03  \\
	    &
[5,5,0]-[5,4,1]  & 1.06  & [5,5,1]-[5,4,2]  & \nodata & [6,1,5]-[5,2,4]
& 1.08 \\
	    &
[6,1,5]-[6,0,6]  & \nodata & [6,2,5]-[6,1,6]  & 1.07: & [6,3,4]-[6,2,5]
& \nodata \\
	    &
[7,2,5]-[6,3,4]  & 1.05 & [7,3,5]-[7,2,6]  & 1.07:  & [8,2,6]-[8,1,7]
& 1.05  \\
	    &
[8,3,6]-[8,2,7]  & \nodata & [9,2,7]-[9,1,8]  & 1.08 & [9,3,6]-[8,4,5]
& \nodata \\
	    &
[9,3,7]-[9,2,8]  & \nodata & [10,4,7]-[10,3,8]  & 1.03 &     &    \\
      &       &       &      &      &        &       \\
(010)-(000) & [6,5,1]-[5,2,4]  &   \nodata & [8,8,1\&0]-[7,7,0\&1]
& \nodata & [9,5,5]-[8,2,6]  &   \nodata \\
	    &
[9,8,1\&2]-[8,7,2\&1] & 1.04 & [10,3,7]-[10,0,10]  & 1.02  &
[10,3,8]-[9,0,9]  & 1.04 \\
	    &
[10,4,7]-[9,1,8]  &  1.06 & [11,3,8]-[10,2,9]  &  1.08 & [11,7,4]-[10,6,5]
&  1.04 \\
	    &
[11,7,5]-[10,6,4]  &  1.03 & [12,4,8]-[11,3,9]  &  \nodata &
[12,6,6]-[11,5,7]  &  1.06 \\
	    &
[13,5,8]-[12,4,9]  &  1.05 & [13,5,9]-[12,4,8]  &  1.07  &
[13,6,7]-[12,5,8]  &  \nodata \\
	    &
[14,6,9]-[13,5,8]  &  1.06 & [15,6,10]-[14,5,9]  &  \nodata &
[16,6,11]-[15,5,10] &  \nodata \\
      &       &       &      &      &        &       \\
(001)-(010) & [0,0,0]-[1,0,1]  & \nodata & [1,1,0]-[1,1,1]  & \nodata
& [1,1,1]-[1,1,0]  & 1.15 \\
	    &
[2,1,2]-[2,1,1]  & \nodata & [2,2,0]-[2,2,1]  & 1.18 & [2,2,0]-[3,0,3]
& \nodata \\
	    &
[2,2,1]-[2,2,0]  & \nodata & [3,0,3]-[2,2,0]  & 1.20 & [3,2,1]-[3,2,2]
& \nodata \\
	    &
[3,2,2]-[3,2,1]  & 1.23:  & [3,2,2]-[4,2,3]  & 1.21  & [4,0,4]-[3,2,1]
& \nodata \\
	    &
[4,2,2]-[4,2,3]  & 1.06 & [5,0,5]-[4,2,2]  & 1.02 & [5,1,4]-[4,3,1]
& \nodata \\
	    &
[6,1,5]-[5,3,2]  & 1.01: & [6,3,3]-[6,3,4]  & 1.02 & [7,1,6]-[6,3,3]
& \nodata \\
	    &
[7,3,4]-[7,3,5]  & 1.09 & [8,1,7]-[7,3,4]  & \nodata & [8,2,6]-[7,4,3]
& \nodata \\
	    &
[8,3,5]-[8,3,6]  & 1.03 &    &   &   &   \\
      &       &       &      &      &        &       \\
(020)-(010) & [5,5,0]-[4,2,3]  &  \nodata & [6,5,2]-[5,2,3]  &  1.01
& [8,8,0\&1]-[7,7,1\&0]  & 1.05 \\
	    &
[10,2,8]-[9,1,9]  & \nodata &  [10,3,8]-[9,0,9]  & \nodata &
[10,4,7]-[9,1,8]  & 1.02 \\
	    &
[10,7,4]-[9,6,3] & 1.03 & [10,7,3]-[9,6,4]  & \nodata & [11,3,8]-[10,2,9]
& \nodata \\ \enddata \end{deluxetable}

\clearpage


\begin{figure}
\epsscale{0.8} \plotone{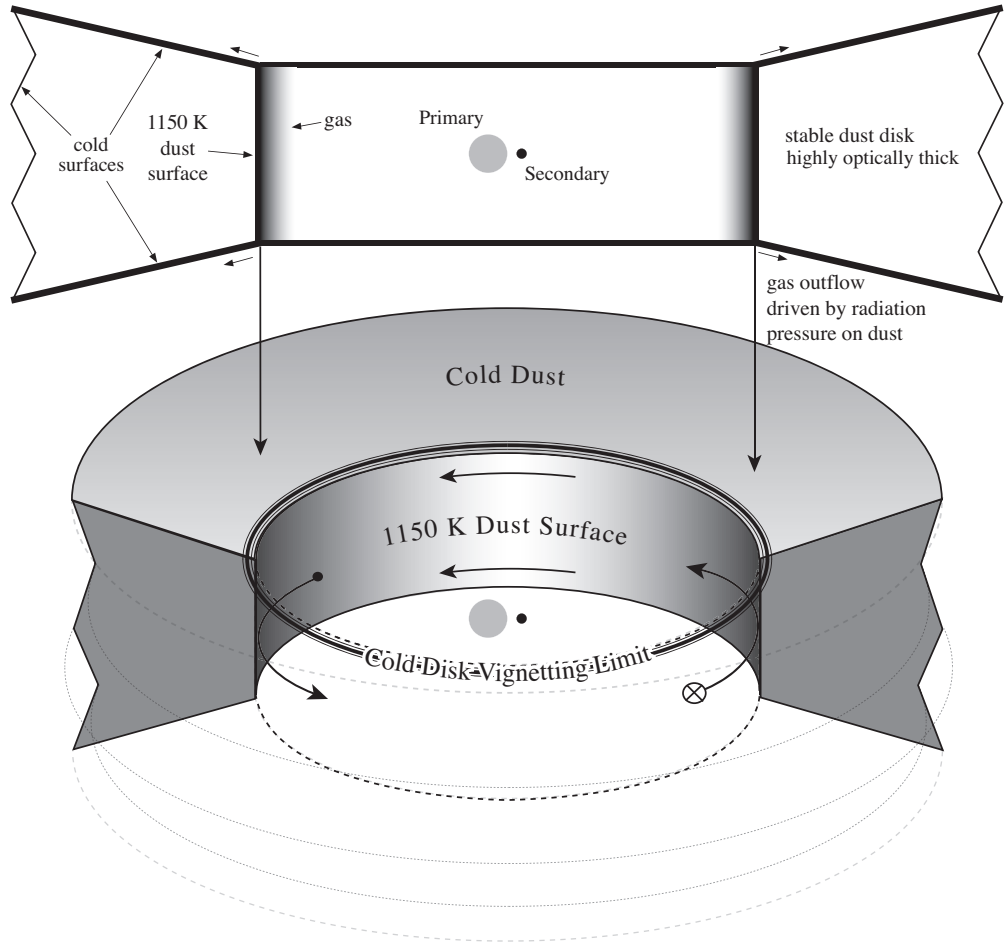} \caption{Cartoon of the `wall' model
for HR 4049 (top) taken from \citet{dominik_et_al_2003}.  Below the Dominik
model the spatially resolved observer's view of the system is shown
in cross section.  The disk is illuminated only from the inside.
The cold disk blocks a large section of the inner 1150 K surface
from view.  The observer sees only that section of the 1150 K 
disk inside the oval labeled ``Cold Disk Vignetting Limit.''} 
\end{figure}


\begin{figure} \epsscale{0.8}
\plotone{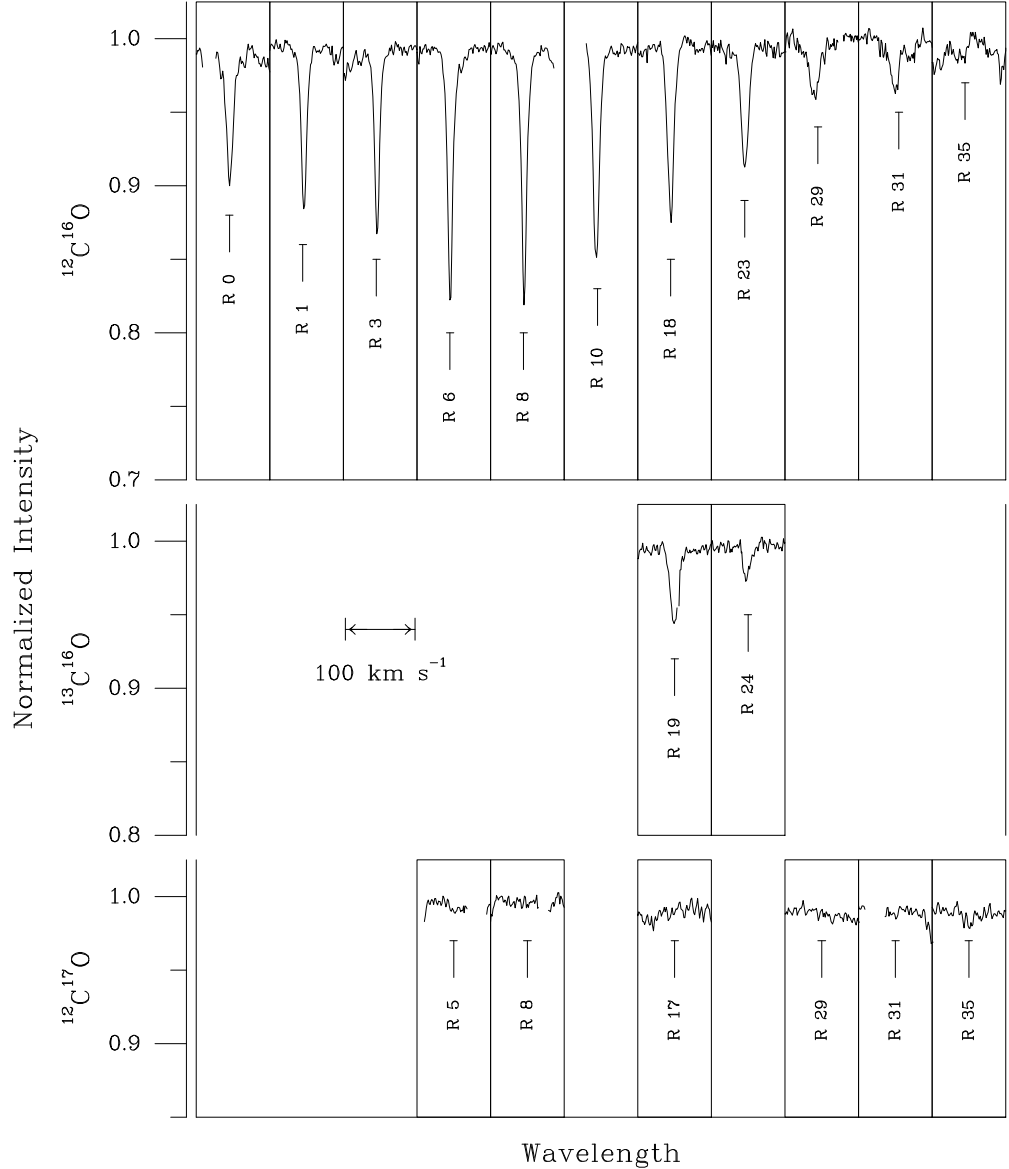} \caption[]{A selection of CO 2-0 R branch lines.
The abscissa consists of $\sim$8 \AA~(1.5 cm$^{-1}$ or $\sim$100 km s$^{-1}$)
increments of spectrum centered on each of the labeled lines.  
Top row shows $^{12}$C$^{16}$O lines, middle row $^{13}$CO 
lines and bottom row C$^{17}$O lines.  The columns aline the rotation quantum 
number J for the isotopic lines to approximately
equal excitation (within J''$\pm$1).  The
spectral region containing $^{13}$CO was not well covered, hence
only a few lines are shown, however, $^{13}$CO lines are clearly
present in the spectrum.  Only limits to the
C$^{17}$O lines are detected.  All CO first overtone lines have
purely absorption profiles.  } \end{figure}


\begin{figure}
\epsscale{0.8} \plotone{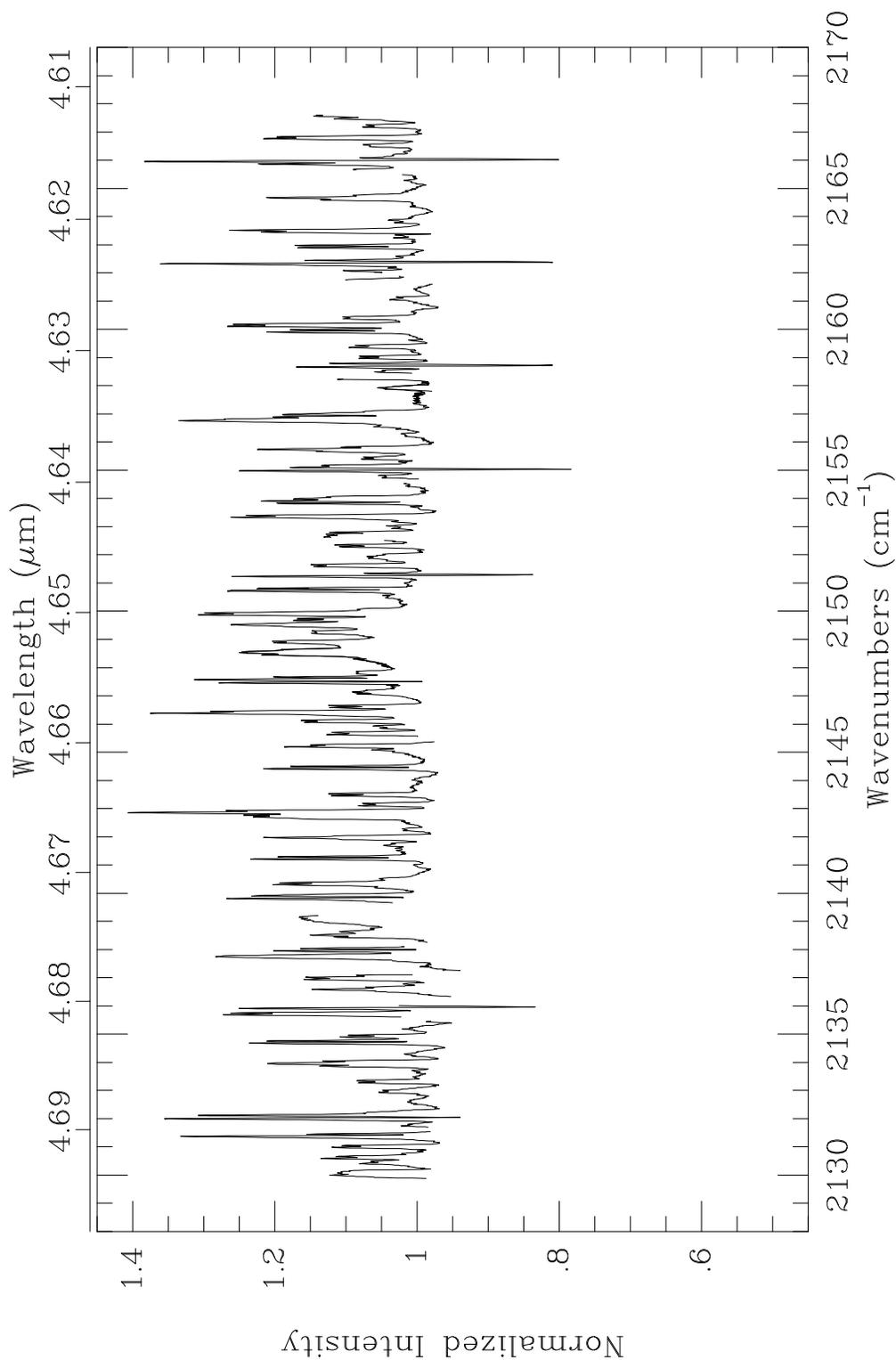} \caption{Overview of the 4.61 --
4.69 $\mu$m spectrum of HR 4049 showing the forest of CO and H$_2$O
emission lines.  Gapped spectral regions indicate
failure to restore the spectrum of HR 4049 due to optically thick
telluric lines.  } \end{figure}


\begin{figure}
\epsscale{0.8} \plotone{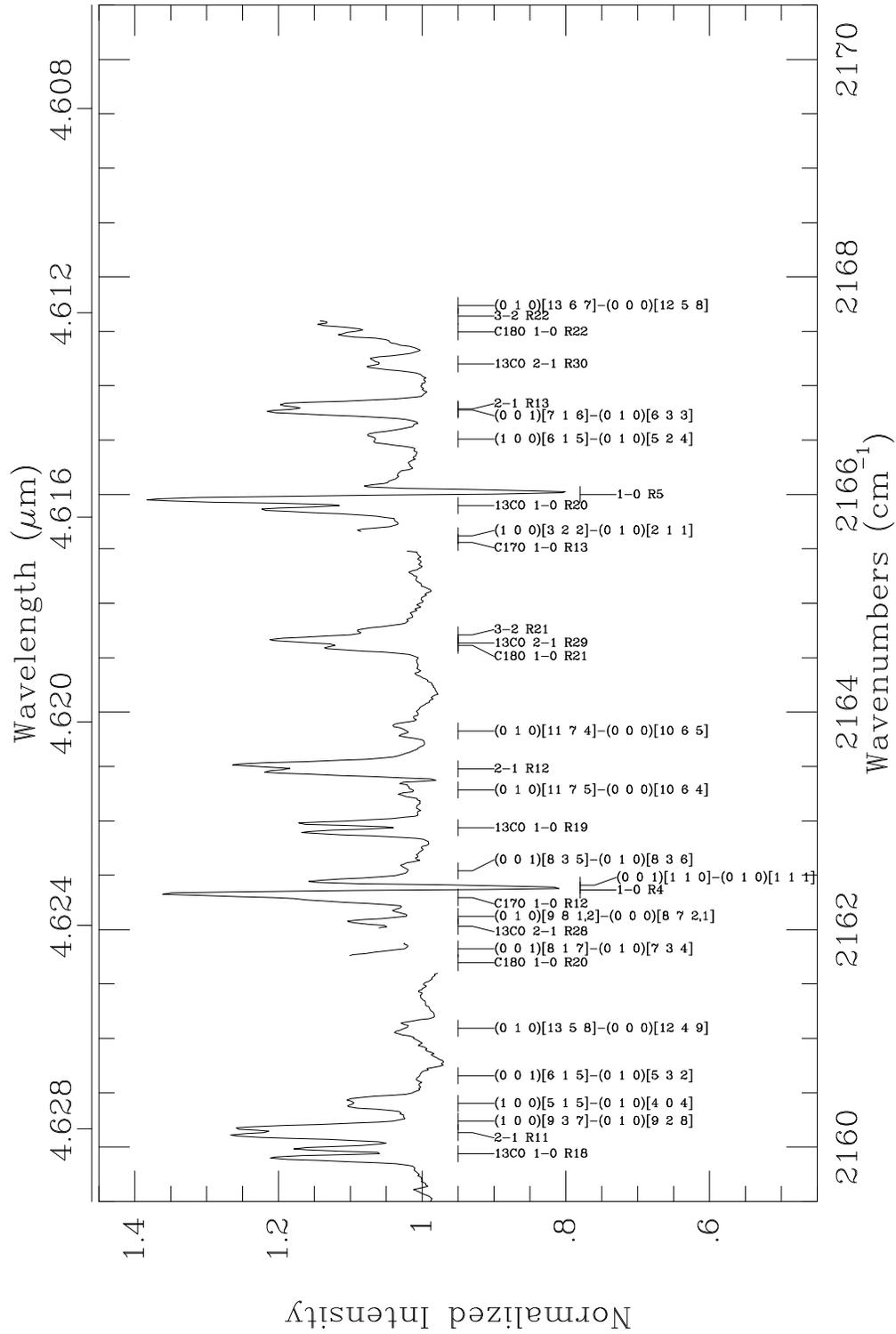} \caption{The 4.62$\mu$m region
spectrum of HR 4049 with line identifications.} \end{figure}


\begin{figure}
\epsscale{0.8} \plotone{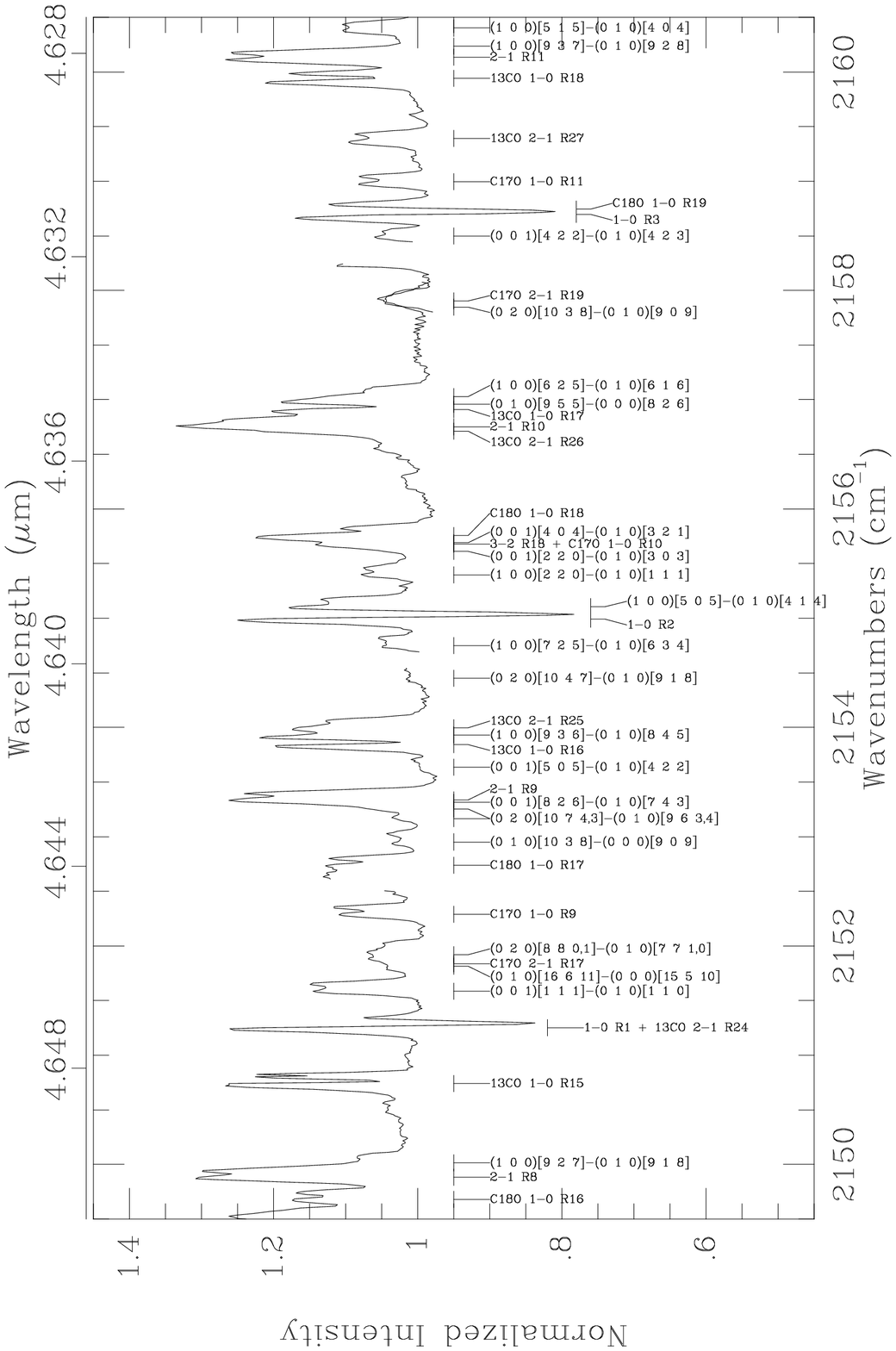} \caption{As per Figure 4 for the
4.64$\mu$m region.} \end{figure}


\begin{figure}
\epsscale{0.8} \plotone{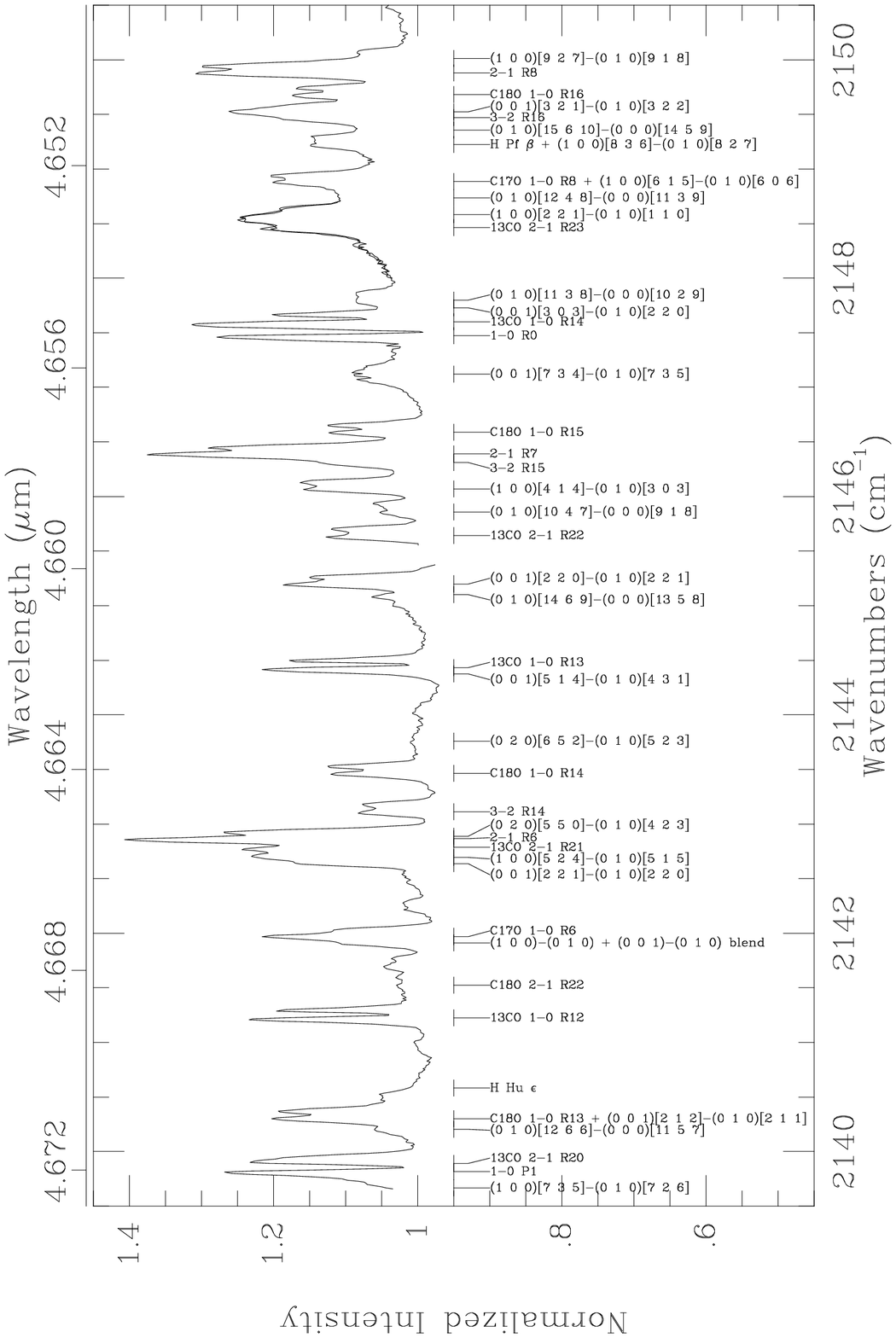} \caption{As per Figure 4 for the
4.66$\mu$m region.} \end{figure}


\begin{figure}
\epsscale{0.8} \plotone{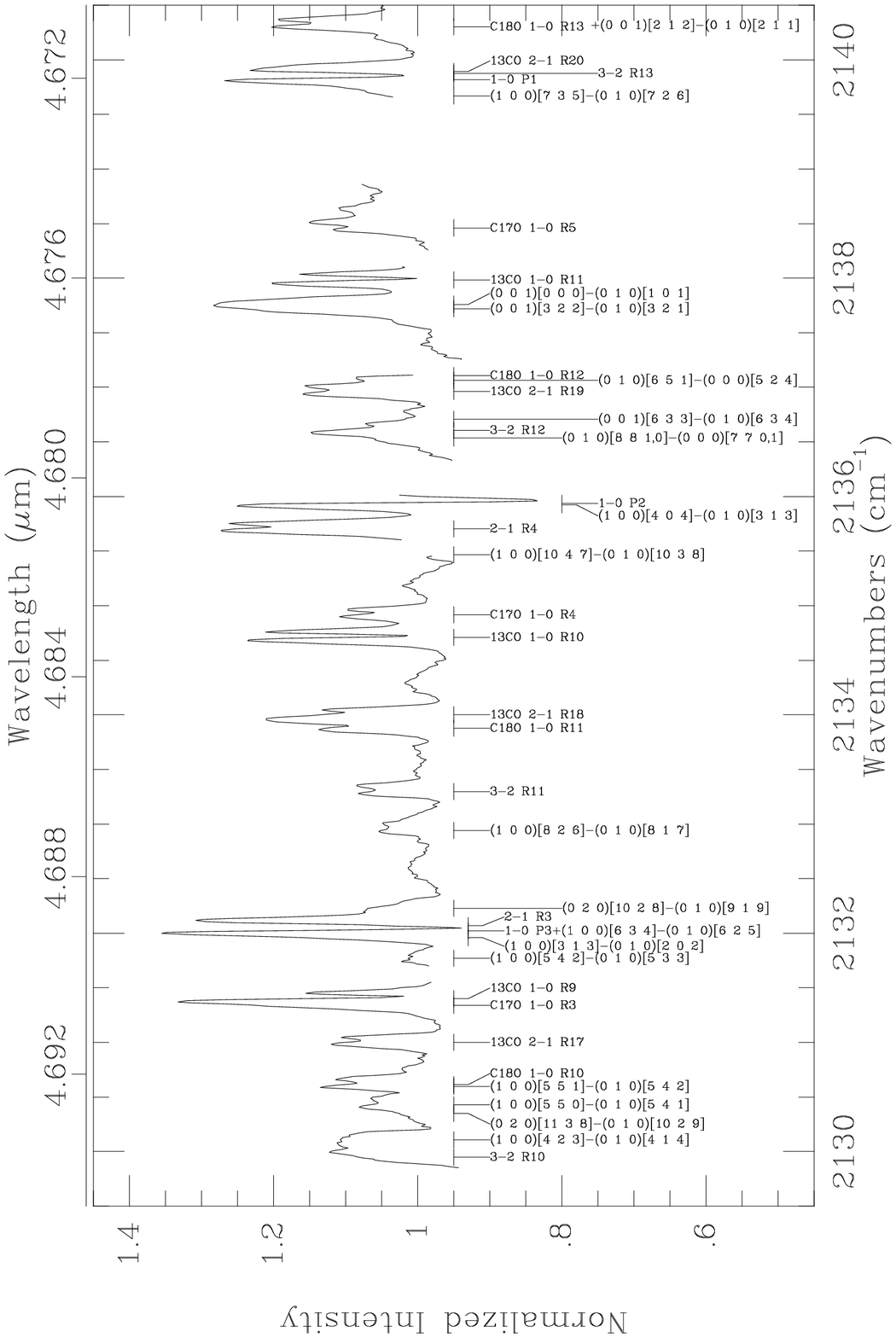} \caption{As per Figure 4 for the
4.68$\mu$m region.} \end{figure}


\begin{figure}
\epsscale{0.8} \plotone{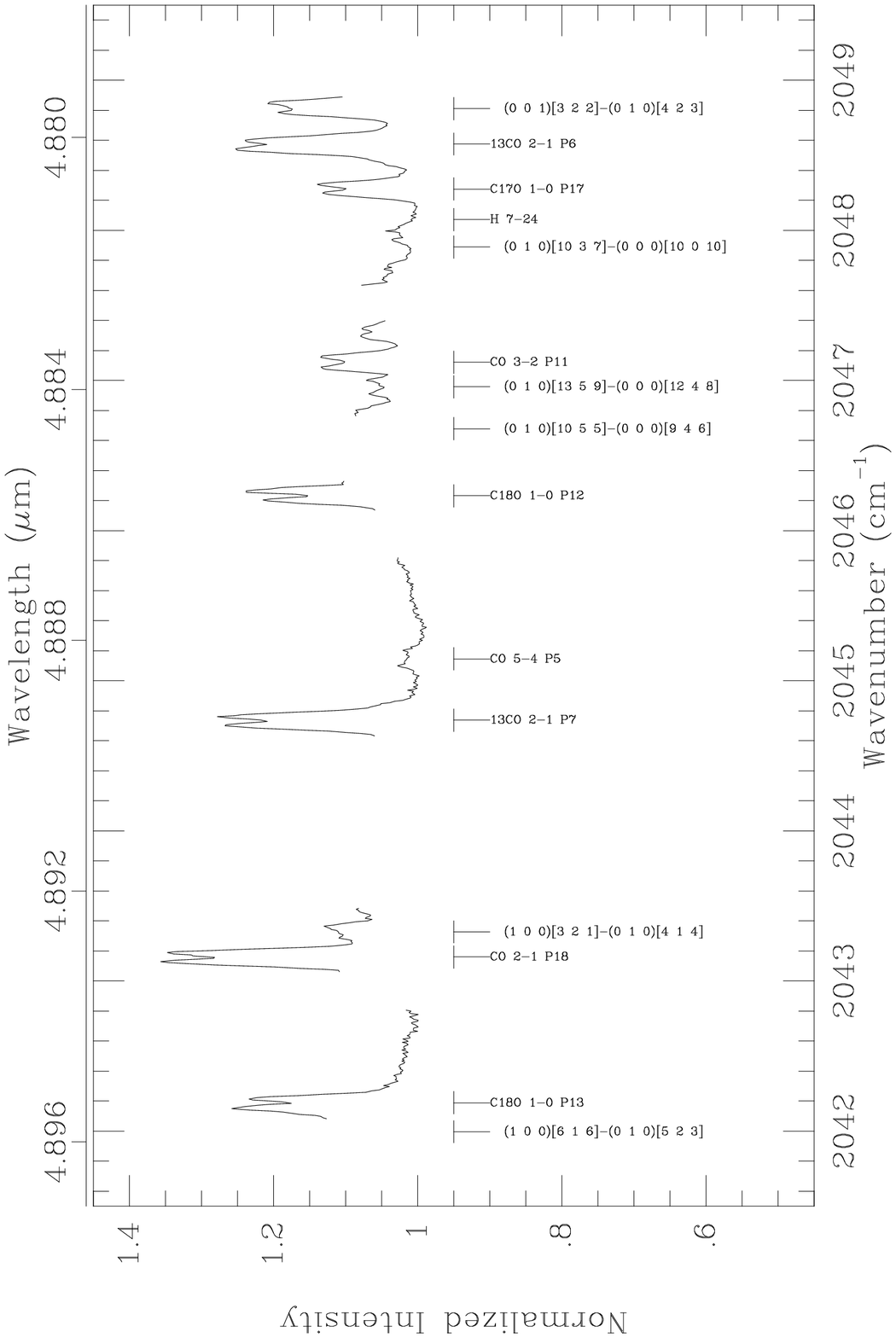} \caption{As per Figure 4 for the
4.89$\mu$m region.} \end{figure}


\begin{figure}
\epsscale{0.8} \plotone{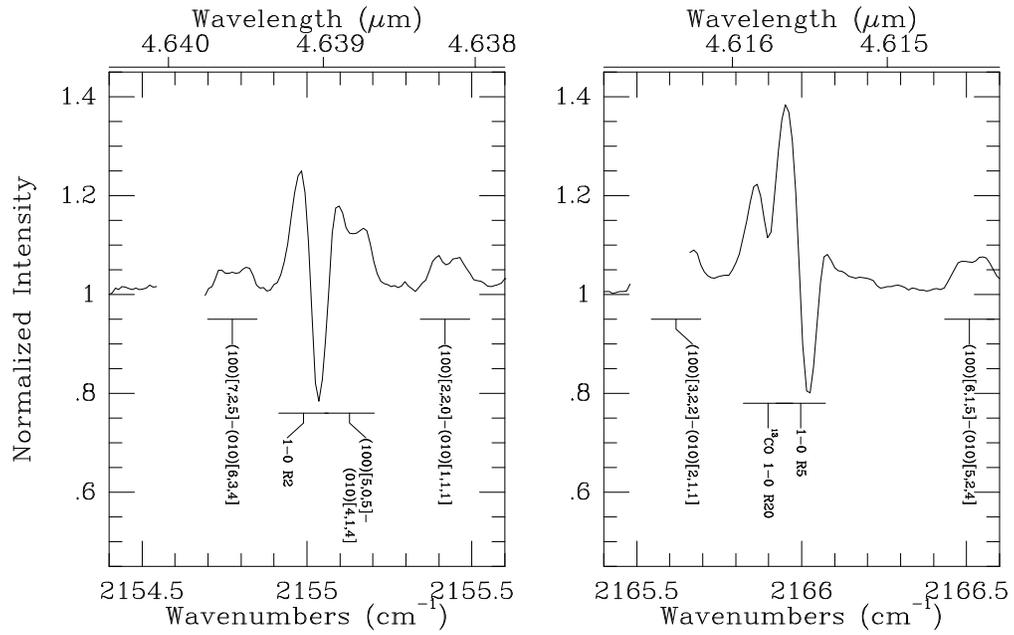} \caption{An enlarged view of the
spectrum shown in Figures 4 and 5 showing the regions surrounding
the $^{12}$C$^{16}$O R 2 line (left) and the R 5 line (right).  The
R 2 line is unblended on the red wing while the R 5 line is unblended
on the blue wing.  Both lines have P-Cygni type profiles.  Higher
excitation CO lines as well as H$_2$O lines shown in this Figure
exhibit typical double peaked emission profiles.  } \end{figure}


\begin{figure}
\epsscale{0.6} \plotone{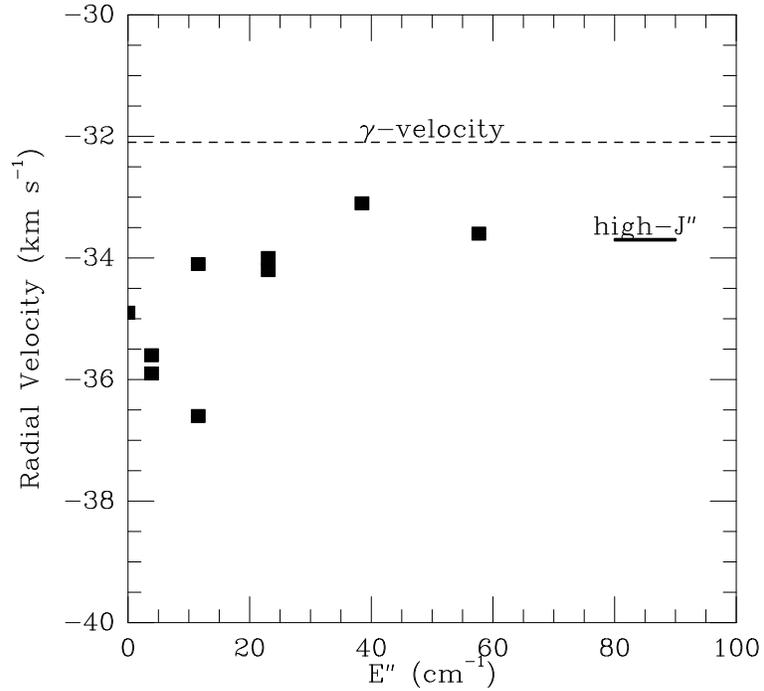} \caption{ Radial velocities of the
absorption component of the CO 1-0 low excitation lines (J''=0
through 5) as a function of excitation energy of the lower level.
A number of these lines are blended with other circumstellar lines.
The bar to the right labeled `\,high-J''\,' is at the value of the
mid-emission absorption for higher excitation lines.  There is a
clear trend for the lowest excitation lines to have a larger outflow.
The dashed line is the binary system $\gamma$-velocity
\citep{bakker_et_al_1998}.  } \end{figure}


\begin{figure} \epsscale{0.5}
\plotone{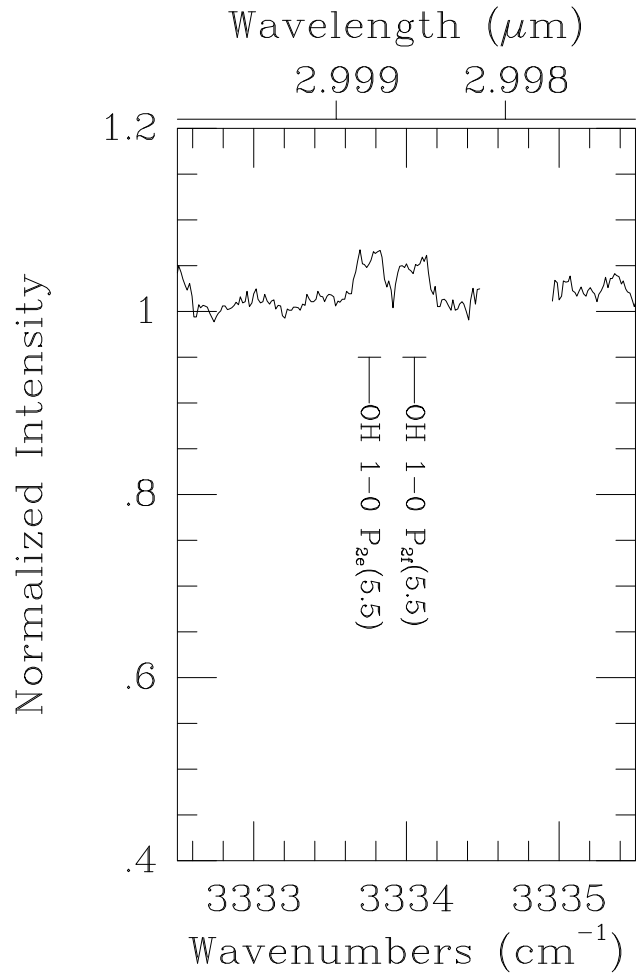} \caption{ The 3$\mu$m region spectrum of HR 4049
showing the OH 1-0 P$_{2f}$5.5 and 1-0 P$_{2e}$5.5 lines.  }
\end{figure}


\begin{figure} \epsscale{1.0} \plotone{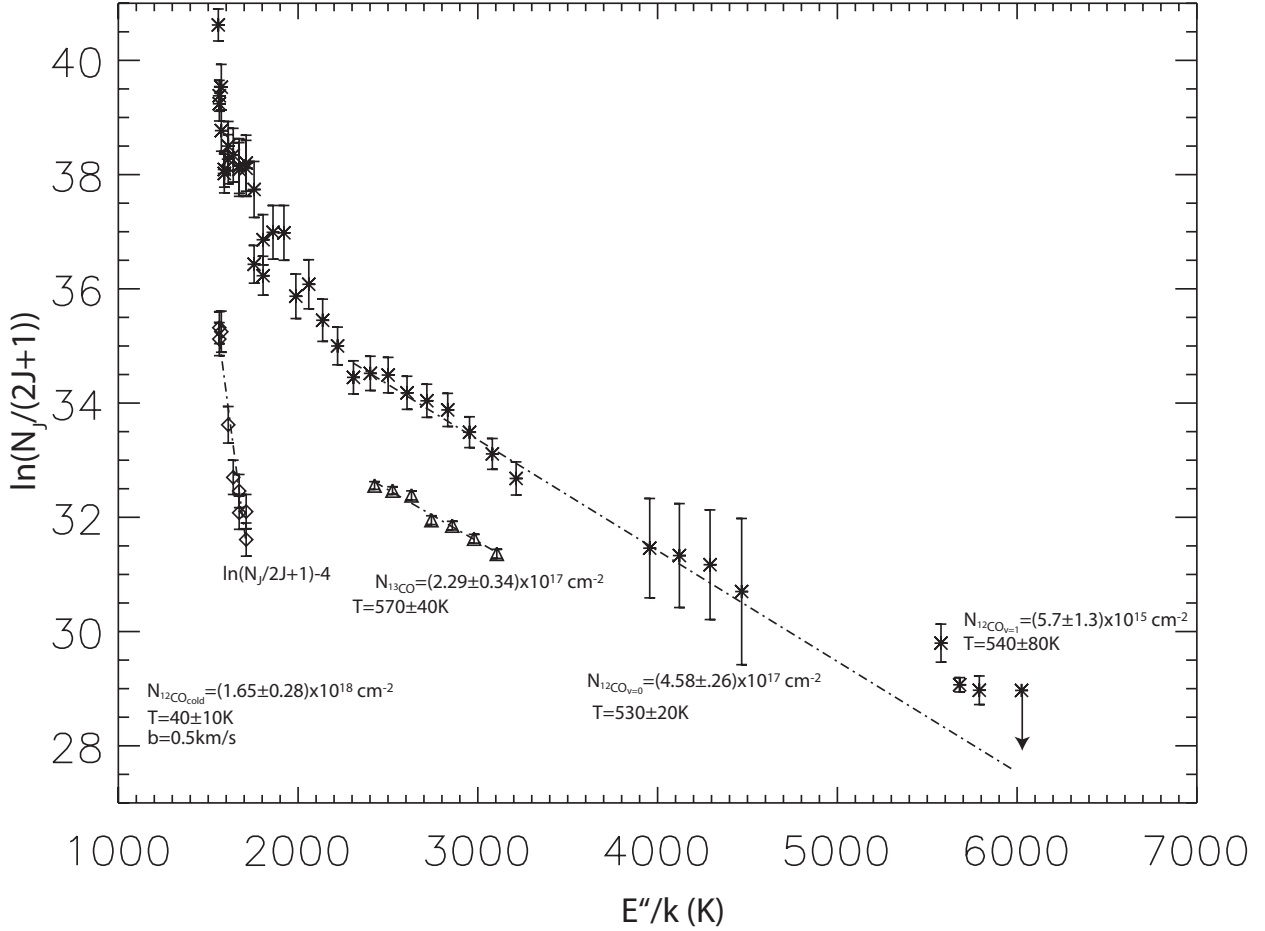} \caption{Boltzman
plot for HR 4049 first overtone CO lines.  Data are shown for the
two isotopic species of CO that were detected in the first overtone
spectra, $^{12}$C$^{16}$O and $^{13}$C$^{16}$O.  The $^{12}$C$^{16}$O
excitation temperature is 40$\pm$10 K the low excitation lines and
530$\pm$20 K for the high excitation 2-0 lines.  The four 3-1
$^{12}$C$^{16}$O lines populate the 5000-6000 K region of the
abscissa with an excitation temperature of 540$\pm$80K, suggesting
a slight departure from vibrational LTE.  However, a fit through
all higher excitation lines remains within the uncertainties and
gives an excitation temperature of 620$\pm$20 K.  The $^{13}$CO
lines have an excitation temperature of 570$\pm$40 K.  } \end{figure}


\begin{figure} \epsscale{0.5}
\plotone{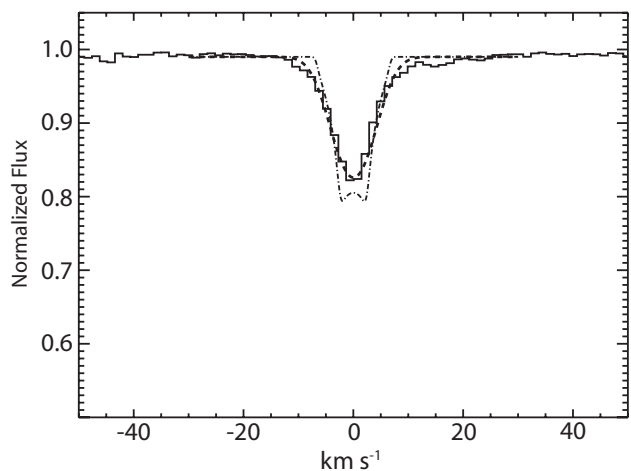} \caption{Synthetic line profiles for the CO first
overtone.  The dot-dash line results from modeling a thin gas layer
on the interior surface of the dust disk (see text).  The dash line
is the same model spectral line convolved to the instrumental
resolution.  The solid line is the observed profile of the CO R6
line. } \end{figure}


\begin{figure} \epsscale{0.5}
\plotone{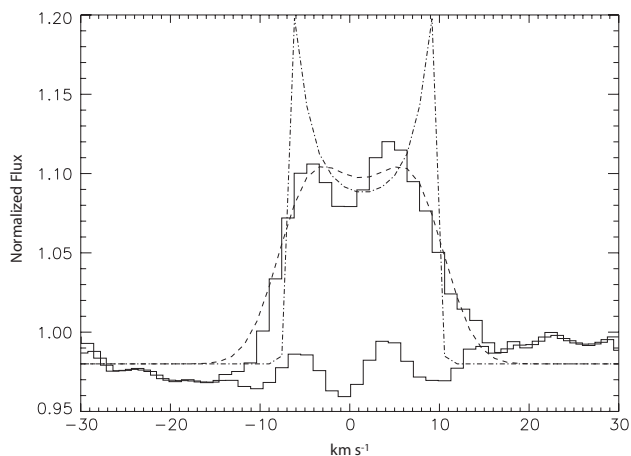} \caption{ A synthetic line profile for the CO
fundamental lines compared to an observed profile.  The dot-dash
line is the synthetic line profile from an emitting zone near the
`cold disk vignetting limit' in Figure 1 (see text).  The dash line
shows this profile convolved to the instrumental resolution.  The
upper solid line is an observed CO line profile and the lower solid
line is the difference between the model and observed line profile.
} \end{figure}


\begin{thebibliography}{}

\bibitem[Aleon et al.(2005)]{aleon_et_al_2005} Aleon, J., Robert,
F., Duprat, J., \& Derenne, S. 2005, Nature, 437, 385

\bibitem[Antoniucci et al.(2005)]{antoniucci_et_al_2005} Antoniucci,
S., Paresce, F., \& Wittkowski, M. 2005, \aap, 429, L1

\bibitem[Ayres \& Wiedemann(1989)]{ayres_wiedemann_1989} 
Ayres, T. R. \& Wiedemann, G. R. 1989, \apj, 338,
1033

\bibitem[Bakker et al.(1996)]{bakker_et_al_1996} Bakker, E. J., van
der Wolf, F. L. A., Lamers, H. J. G. L. M., Gulliver, A.  F., Ferlet,
R., \& Vidal-Madjar, A.  1996, \aap, 306, 924

\bibitem[Bakker et al.(1998)]{bakker_et_al_1998} Bakker, E. J.,
Lambert, D. L., Van Winckel, H., McCarthy, J. K., Waelkens, C., \&
Gonzalez, G.  1998, \aap, 336, 263

\bibitem[Beintema et al.(1996)]{beintema_et_al_1996} Beintema, D.
A., van den Ancker, M. E., Molster, F. J., Waters, L. B.  F.  M.,
Tielens, A. G. G. M., Waelkens, C., de Jong, T., de Graauw, T.,
Justtanont, K., Yamamura, I., Heras, A., Lahuis, F., \& Salama, A.
1996, \aap, 315, L369

\bibitem[Bergeron et al.(1992)]{bergeron_et_al_1992} Bergeron, P.,
Saffer, R. A., \& Liebert, J. 1992, \apj, 394, 228

\bibitem[Bond(1991)]{bond_1991} Bond, H. E. in Proc. IAU Symposium
145, ``Stellar Evolution: The Photospheric Abundance Connection,''
eds. G. Michaud \& A. Tutekov, Reidel:Dordrecht, p. 341

\bibitem[Brittain et al.(2005)]{brittain_2005} Brittain, S.~D.,
Rettig, T.~W., Simon, T., \& Kulesa, C. 2005, \apj, 626, 283

\bibitem[Cami \& Yamamura(2001)]{cami_yamamura_2001} Cami, J. \&
Yamamura, I. 2001, \aap, 367, L1

\bibitem[Clayton et al.(2005)]{clayton_et_al_2005} Clayton, G. C.,
Herwig, F., Geballe, T. R., Asplund, M., Tenebaum, E. D., Engelbracht,
C. W., \& Gordon, K. D. 2005, \apj, 623, L141

\bibitem[Clayton et al.(2007)]{clayton_et_al_2007} Clayton, G. C.,
Geballe, T. R., Herwig, F., Fray, C., Asplund, M. 2007, preprint.

\bibitem[Dominik et al.(2003)]{dominik_et_al_2003} Dominik, C.,
Dullemond, C. P., Cami, J., \& van Winckel, H. 2003, \aap, 397, 595

\bibitem[De Ruyter et al.(2006)]{de_ruyter_et_al_06} De Ruyter, S.,
Van Winckel, H., Maas, T., Lloyd Evans, T., Waters, L.B.F.M., \&
Dejonghe, H. 2006, \aap, 448, 641

\bibitem[Geballe et al.(1989)]{geballe_et_al_1989} Geballe, T. R.,
Noll, K. S., Whittet, D. C. B., \& Waters, L. B. F. M.  1989, \apj,
340, L29

\bibitem[Glassgold et al.(2004)]{glassgold_et_al_2004}
Glassgold, A. E., Najita, J., \& Igea, J. 2004, \apj, 615, 972

\bibitem[Grevesse et al.(1991)]{grevesse_et_al_1991} Grevesse, N,
Lambert, D. L., Sauval, A. J., van Dishoek, E. F., Farmer, C.  B.,
\& Norton, R. H. 1991, \aap, 242, 488

\bibitem[Guillois et al.(1999)]{guillois_et_al_1999} Guillois, O.,
Ledoux, G., \& Reynaud, C. 1999, \apj, 521, L133

\bibitem[Hartmann et al.(1998)]{hartmann_et_al_1998} Hartmann, L.,
Calvet, N., Gullbring, E., \& D'Alessio, P. 1998, \apj, 495, 385

\bibitem[Hinkle \& Barnes(1979)]{hinkle_barnes79} Hinkle, K. H. \&
Barnes, T. 1979, \apj, 227, 923

\bibitem[Hinkle et al.(1998)]{hinkle_et_al_98} Hinkle, K. H.,
Cuberly, R., Gaughan, N., Heynssens, J., Joyce, R., Ridgway, S.,
Schmitt, P., \& Simmons, J. E. 1998, Proc. SPIE, 3354, 810

\bibitem[Hinkle et al.(2000)]{hinkle_et_al_00} Hinkle, K. H., Joyce,
R. R., Sharp, N., \& Valenti, J. A. 2000, Proc. SPIE, 4008, 720

\bibitem[Hinkle et al.(2003)]{hinkle_et_al_03} Hinkle, K. H., Blum,
R., Joyce, R. R., Ridgway, S. T., Rodgers, B., Sharp, N., Smith,
V., Valenti, J., van der Bliek, N.,\& Winge, C. 2003, Proc. SPIE,
4834, 353

\bibitem[Howell et al.(2006)]{howell_et_al_2006} Howell, S. B.,
Brinkworth, C., Hoard, D. W., Wachter, S., Harrison, T., Chun, H.,
Thomas, B., Stefaniak, L., Ciardi, D. R., Szkody, P., van Belle,
G. 2006, \apj, 646, L65

\bibitem[Joyce(1992)]{joyce_92} Joyce, R. R. 1992, in ASP Conf.
Ser. 23, Astronomical CCD Observing and Reduction Techniques, ed.
S. Howell (San Francisco: ASP), p. 258

\bibitem[Kurucz(1979)]{kurucz_1979} Kurucz, R. L. 1979, \apjs, 40,
1

\bibitem[Lambert et al.(1986)]{lambert_et_al_1986} Lambert, D. L.,
Gustafsson, B., Eriksson, K., \& Hinkle, K. H. 1986, \apjs, 62, 373

\bibitem[Lambert(1988)]{lambert_1988} Lambert, D. L. 1988 in Evolution
of Peculiar Red Giant Stars, IAU Colloq. 106, eds. H. R. Johnson
\& B. M. Zuckerman (Cambridge: Cambridge Univ. Press), p. 101

\bibitem[Lambert et al.(1988)]{lambert_et_al_1988} Lambert, D. L.,
Hinkle, K. H., \& Luck, R. E. 1988, \apj, 333, 917

\bibitem[Lugaro et al.(2005)]{lugaro_et_al_2004} Lugaro, M., Pols,
O., Karakas, A. I., \& Tout, C. A. 2005, Nuclear Physics A, 758,
725

\bibitem[Mathis \& Lamers(1992)]{mathis_lamers_1992} Mathis, J. S.
\& Lamers, H. J. G. L. M. 1992, \aap, 259, L39

\bibitem[McClatchey et al.(1973)]{mcclatchey} McClatchey, R. A.,
Benedict, W. S., Clough, S. A., Burch, D. E., Calfee, R. F., Fox,
K., Rothman, L. S., \& Garing, J. S. 1973, ``AFCRL Atmospheric
Absorption Line Parameters Compilation'', Environmental Research
Papers, 434 (Bedford, Mass.: Air Force Cambridge Research Laboratories)

\bibitem[Najita et al.(1996)]{najita_et_al_1996}
Najita, J., Carr, J. S., Glassgold, A. E., Shu, F. H., Tokunaga, A. T. 1996,
\apj, 462, 919

\bibitem[Netzer \& Elitzur(1993)]{netzer_elitzur_1993} Netzer, N.
\& Elitzur, M. 1993, \apj, 410, 701

\bibitem[Rasio \& Livio(1996)]{rasio_livio_1996} Rasio, F. A. \&
Livio, M. 1996, \apj, 471, 366

\bibitem[Smith et al.(2002)]{smith02} Smith, V. V., Hinkle, K. H.,
Cunha, K., Plez, B., Lambert, D. L., Pilachowski, C. A., Barbuy,
B., Melendez, J., Balachandran, S., Bessell, M. S., Geisler, D. P.,
Hesser, J. E., \& Winge, C. 2002, \aj, 124, 3241

\bibitem[Spitzer(1978)]{spitzer_1978} Spitzer, L. 1978, Physical
Processes in the Interstellar Medium (New York, Wiley)

\bibitem[Taam \& Sandquist(2000)]{taam_sandquist_2000} Taam, R. E.
\& Sandquist, E. L. 2000, \araa, 38, 113

\bibitem[Takada-Hidai(1990)]{takada_hidai_90} Takada-Hidai, M. 1990,
\pasp, 102, 139

\bibitem[Takeuchi \& Artymowicz(2001)]{takeuchi_artymowicz_2001}
Takeuchi, T. \& Artymowicz, P. 2001, \apj, 557, 990

\bibitem[Takeuchi \& Lin(2002)]{takeuchi_lin_2002} Takeuchi, T. \&
Lin, D. N. C. 2002, \apj, 581, 1344

\bibitem[Takeuchi \& Lin(2003)]{takeuchi_lin_2003} Takeuchi, T. \&
Lin, D. N. C. 2003, \apj, 593, 524

\bibitem[van Winckel et al.(1995)]{van_winckel_et_al_95} van Winckel,
H., Waelkens, C., \& Waters, L. B. F. M. 1995, \aap, 293, L25

\bibitem[Venn \& Lambert(1990)]{venn_lambert_1990} Venn, K. A. \&
Lambert, D. L. 1990, \apj, 363, 234

\bibitem[Waters et al.(1989)]{waters_et_al_1989} Waters, L. B. F.
M., Lamers, H. J. G. L. M., Snow, T. P., Mathlener, E., Trams, N.
R., van Hoof, P. A. M., Waelkens, C., Seab, C. G., \& Stanga, R.
1989, \aap, 211, 208


\bibitem[Waters et al.(1992)]{waters_et_al_1992} Waters, L. B. F.
M., Trams, N. R., \& Waelkens, C. 1992, \aap, 262, L37

\bibitem[Waters et al.(1998)]{waters_et_al_1998} Waters, L. B. F.
M., et al. 1998, Nature, 391, 868

\bibitem[Waelkens et al.(1991a)]{waelkens_et_al_91a} Waelkens, C.,
Lamers, H. J. G. L. M., Walters, L. B. F. M., Rufener, F., Trams,
N. R., LeBertre, T., Ferlet, R., \& Vidal-Madjar, A. 1991a, \aap,
242, 433

\bibitem[Waelkens et al.(1991b)]{waelkens_et_al_91b} Waelkens, C.,
Van Winckel, H., Bogaert, E., \& Trams, N. R. 1991b, \aap, 251, 495


\bibitem[Waelkens et al.(1996)]{waelkens_et_al_96} Waelkens, C.,
Van Winckel, H., Waters, L.B.F.M., \& Bakker, E.J. 1996, \aap, 314,
L17

\end{thebibliography}
\end{document}